\documentclass{iopjournal}

\usepackage{tablefootnote}

\begin{document}

\articletype{Topical Review} 

\title{Electromagnetic Probes of the Supernova Engine}

\author{Chris L. Fryer$^1$\orcid{0000-0003-2624-0056}}

\affil{$^1$Center for Nonlinear Studies, Los Alamos National Laboratory, Los Alamos, NM 87545 USA}


\email{fryer@lanl.gov}


\begin{abstract}

Explosions from stellar collapse are important in nearly all aspects of astronomy:  formation of compact remnants (including the potential seeds of supermassive black holes), the production of compact binaries the produce X-ray binaries and radio pulsars, the origin of many of the heavy elements (a critical aspect of galactic chemical evolution), and as standard candles to probe the early universe (including the nature of star formation).  Because these explosions occur in extreme conditions, if we can understand them, they can be used to probe fundamental physics:  matter at extreme densities (exceeding nuclear densities, quark formation), neutrino physics (including neutrino oscillations) and, in black-hole forming systems, the nature of relativity.  Gravitational waves and neutrinos provide the most direct probe of stellar collapse, but these diagnostics require nearby events.  Electromagnetic probes, from shock breakout to observations of supernova remnants, can be much more common and provide complementary diagnostics of the explosive engines behind stellar collapse.  In this review, we review these electromagnetic diagnostics, the physics behind them and the theory and modeling work we require to take advantage of these probes.

\end{abstract}

\section{The Diverse Probes Observed Through Electromagnetic Waves}

Despite the critical importance of core-collapse supernovae to a broad range of astrophysics and physics disiciplines, the exact nature of the explosion mechanism remains a matter of, sometimes heated, debate.  Although the neutrino-driven mechanism (with and without convection above the proto-neutron star to enhance the neutrino heating) has garnered much success in the past couple decades demonstrating that it is behind most normal supernovae, there is also evidence that jet/disk engines explain at least a subset of observed transients, including broad-line supernovae, those with and without associated gamma-ray bursts~\cite{2025ApJ...986..185F}.  Energy deposition from magnetars may drive an explosion or, more-likely, dominate the observed electromagnetic wave (EM) emission~\cite{2010ApJ...719L.204W,2017ApJ...841...14M}.  We can use the rapidly-growing set of core-collapse supernova observations to distinguish these different engines and, ultimately, characterize their nature.

In supernovae (SNe), thermal neutrinos and gravitational waves are the most direct probes of the SN engine.  Both thermal neutrinos and gravitational waves are produced in the engine itself:  the collapsing core, the proto-neutron star, and the convective material above the proto-neutron star.  Although these observations are the most powerful probes of the supernova engine, they are difficult to make and are likely to be limited to the Milky Way events for the foreseeable future~\footnote{Although we note that, if jet/disk engines are common, the rapid rotation required would lead to gravitational wave singals that could be detected by next-generation gravitational-wave detectors out to the Virgo cluster.  Ultimately, this will provide a strong constraint on the fraction of supernovae produced by rapidly-spinning systems.}.

\begin{figure}
    \centering
    \includegraphics[width=5.5in]{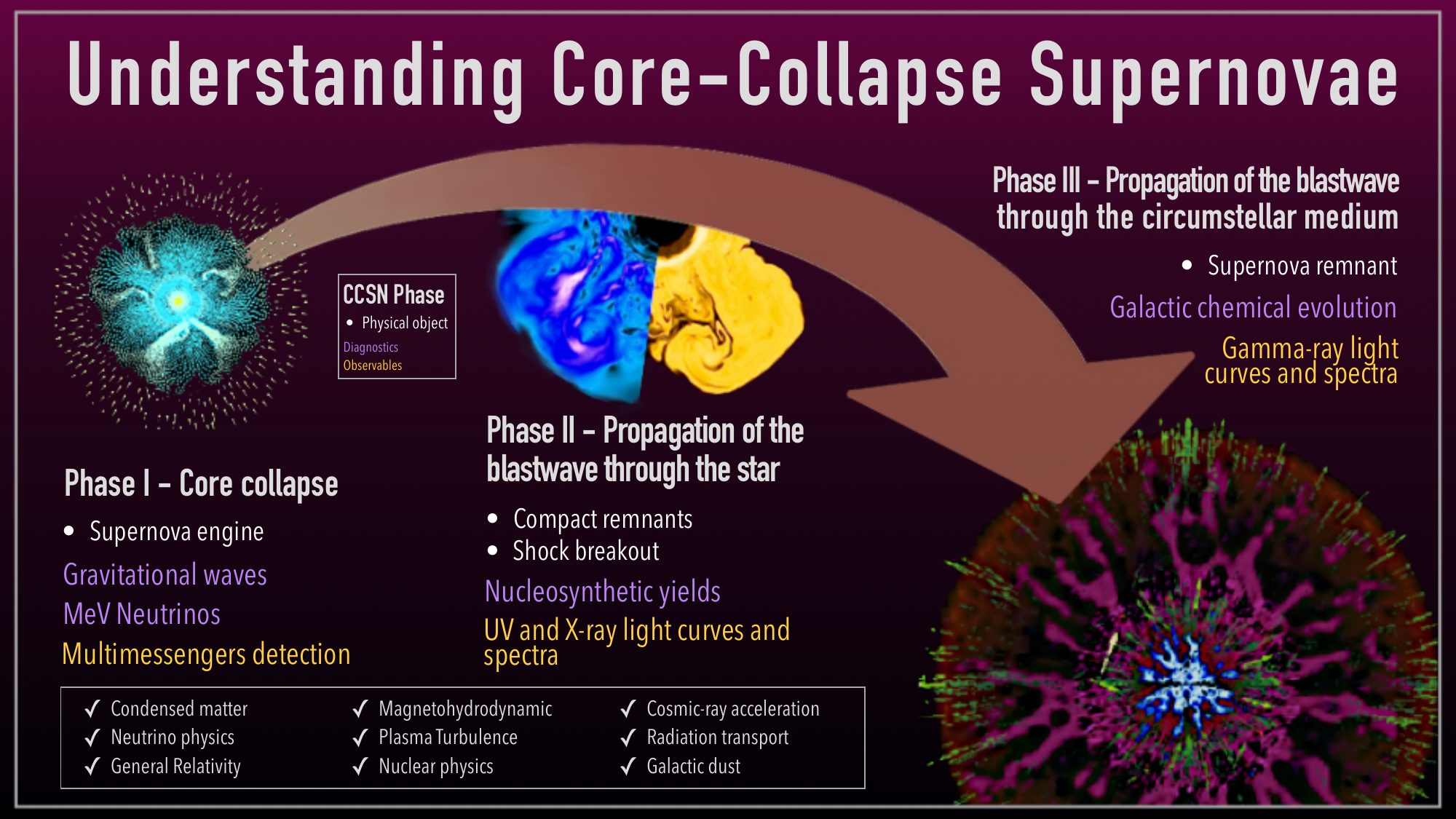}
    \caption{The broad set of observations used to probe the supernova engine.  To do these studies with a full understanding of the uncertainties requires combining multiple studies using a variety of codes implementing different physics in different regimes.  Gravitational waves and neutrino emission peak near, or shortly after, core bounce ($t_0$).  Shock breakout occurs $t_0$+30-300\,s (Type Ib/c supernovae) or $t_0$+$\sim 10,000$\,s (Type IIP supernovae).  Depending on the amount of asymmetry-driven mixing, MeV Gamma-Ray can rise shortly thereafter (peaking anywhere between 5\,d to beyond 100\,d.)  In UVOIR bands, shock interactions and shock cooling dominates after shock breakout and peaks at roughly 10-20\,d.  This is Figure 1 from \cite{2023ApJ...956...19F}.}
    \label{fig:diag}
\end{figure}

We observe supernovae from a range of diagnostics at photons across the entire EM spectrum.  All of these observations contribute to our understanding of these complex explosions (Figure~\ref{fig:diag} summarizes these observations).  These observations include everything from neutrino and gravitational wave emission during the collapse and launch of the explosion to studies many years later of the ejecta and compact remnants.  Many of these diagnostics~\footnote{The different astrophysical observations correspond to diagnostics in a laboratory experiment.  Laboratory experiments leverage all these diagnostics to probe the experiment. Some use ``Messengers'' to refer to different particles from an event, whether or not it is diagnosing an event.  As this paper focuses on understanding an event, we will use the laboratory definition of diagnostic, using messenger and diagnostic interchangeably.} or ``messengers'' are observed by astronomers through EM emission.  Here we review these EM observations.

EM observations probe a wide range of stages in the supernovae and, depending on the exact nature of the probe, can be observed at a much higher rate than neutrinos and gravitational waves.  Table~\ref{tab:diagnostics} shows the list of diagnostics that help us probe the supernova engine, the primary photon energies used in this probe, the energy source for these probes, and the leading explosion properties constrained by the probes.  We also list the number or rate of well-studied (or can be well-studied)~\footnote{Well-studied is a subjective concept, and the number/rate will differ based on its definition.  But we will provide a range based on this definition.} observations for each of these.  Given that we have only detected the engine neutrinos from a single supernova event (SN 1987A~\cite{1987PhRvL..58.1490H,1987PhRvL..58.1494B}) and have yet to detect gravitational waves from a supernova, EM observations are critical to our understanding of supernovae.  As the number of neutrino or graviational wave observations grow, we can tie them to our EM observations, building on our understanding of these cosmic explosions.

But inferring properties of the supernova engine from EM observations is limited by the complexity of these observations.  The emission can come from different regions in the explosion and is powered by different energy sources.  For example, high energy neutrinos and cosmic rays likely arise from interactions of the supernova blastwave with the circumstellar medium.  In these interactions, high energy particles are produced through fermi acceleration across the shock.  However, the uncertainties in the particle acceleration mechanism make it extremely difficult to learn anything about the supernova from these observations.  Indeed, such observations are a better probe of the particle acceleration mechanism and its underlying plasma physics.  As they are not really a strong probe of supernovae themselves, we will not discuss these observations in this paper.

Supernova progenitor observations also place strong constraints on our understanding of supernovae.  Thanks to a number of excellent observing programs leveraging survey observatories such as the Hubble Space Telescope, scientists have built up a large sample of supernova progenitors just prior to collapse.  These observations have placed strong constraints on which stars produce supernova explosions. We will not discuss this constraint further, instead referring the reader to an excellent review by~\cite{2009ARA&A..47...63S}.  

Instead, this review will focus on prompt emission from the ejecta from supernova engine and the remnants of these explosions:  shock breakout, light curves, spectra, compact remnants and the ejecta remnants.  To understand these probes, we must first understand the sources of the photons by which we observe them.  EM photons are produced by a number of mechanisms:  thermalized emission (from internal shocks), shocks (during and after shock breakout), radioactive decay, and synchrotron emission.  These emission processes occur throughout the explosion driving emission from the initial emergence of the SN blastwave (shock breakout) until the supernova remnant fades as it slows.  Disentangling the different sources is further complicated by the processes that alter these signals in their transport from their source to astrophysical observatories.

In this paper, we review the different SN sources of EM emission (Section~\ref{sec:sources}).  These different emission mechanisms are active in a variety of astrophysical observations and can probe different aspects of the supernova explosion.  We discuss these observations ins section~\ref{sec:observations}.  The complexity of the EM emission mechanisms requires a broad range of multi-physics models.  In Section~\ref{sec:models}, we review the required theoretical and experimental work needed to model EM emission and infer properties of the SN engine from EM observations.  

\begin{table}
\caption{Different EM Diagnostics:  Energy Sources include"  Therm $\equiv$ emission from thermalization in the star, Shock $\equiv$ Shock-heating (shocks near/after shock breakout), Decay $\equiv$ Radioactive Decay, Int $\equiv$ internal source (pulsar, magnetar, accretion), Properties probed include:  Prog $\equiv$ Progenitor Properties, M$_{rem}$ $\equiv$ compact remnant mass and Comp, Struc, M$_{eje}$, E $\equiv$ composition, explosion structure/asymmetry, ejecta mass and total energy respectively (strong probes are in bold-face).}
\centering
\begin{tabular}{l c c c c c}
\hline
Diagnostic & Primary Photon Energies & Physics Sources & Properties Probed & \#/Rate \\
\hline
Progenitor & UVOIR & Therm & {\bf Prog} & $>40$~\cite{2009ARA&A..47...63S} \\
Shock Breakout & UV, X-ray & Shocks &  Prog, M$_{exp}$, E$_{exp}$ & $5+$\tablefootnote{This number should increase 2 orders of magnitude with Einstein Probe and Ultraviolet Explorer~\cite{2026arXiv260300820F}} \\
Light Curves & UVOIR & Therm, Shock, Decay, Int & Prog, M$_{eje}$, E, Comp & $100-10,000$ \\
Spectra & UVOIR & Therm, Shock, Decay, Int & Prog, Comp {\bf E} & $100-1,000$ \\
Compact Rem. & Radio, X-ray, orbits & Int & M$_{rem}$ & $30-100$\\
Ejecta Rem. & Radio, X-ray, $\gamma$-ray & Shock, Decay & {\bf Struc }, M, E, Comp & $10-200$~\cite{2012AdSpR..49.1313F} \\
High-E part. & $\gamma-$rays, $\nu$s & Shock & N/A & $\sim 1$ \\
\hline
\end{tabular}

\label{tab:diagnostics}
\end{table}
\section{Sources}
\label{sec:sources}

\begin{figure}
    \centering
    \includegraphics[width=5.5in]{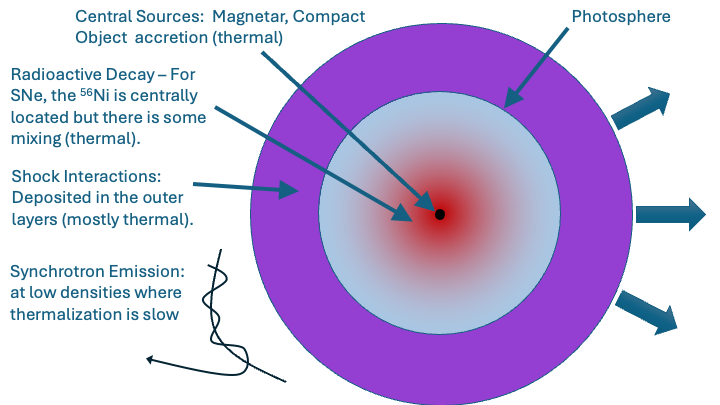}
    \caption{EM sources powering astrophysical transients:  radioactive decay releasing photons and charged particles (primarily $^{56}$Ni for supernovae) that deposit their energy through scattering and absorption, thermal shocks occuring as the blastwave moves through the star and material immediately surrounding the supernova progenitor, synchrotron emission produced in low-density shocks (dominant source producing gamma-ray bursts), and energy sources near the newly-formed compact object (sustained/fallback accretion, magnetars).}
    \label{fig:emsources}
\end{figure}

Emission in astrophysical transients can be powered by and emit through a number of sources and processes including compact remnant sources (magnetars, fallback accretion), radioactive decay, shocks, and nonthermal particle acceleration in shocks.  These different sources are depicted in Figure~\ref{fig:emsources}.  If the source lies within the photosphere, the position where the optical depth of the photons is a few (we will discuss this further in section~\ref{sec:LC}), the deposited energy is likely to thermalize.  Energy sources depositing energy beyond the photosphere (e.g. external shocks, nonthermal particle acceleration) will produce spectra that deviate beyond a thermal blackbody.

\subsection{Radioactive Decay}
\label{sec:source-decay}

One of the best-known power sources for astrophysical transients is that of radioactive decay.  Radioactive isotopes decay, emitting $\gamma$-rays and charged particles that, in turn, thermalize, depositing their energy into the ejecta.  For example, $^{56}$Ni is produced in the engines behind stellar collapse.  It decays to $^{56}$Co with a half-life of 5\,d which, in turn, decays to $^{56}$Fe with a half-life of 77\,d.  In both decays, MeV $\gamma$-rays and positrons are emitted that then down-scatter (in the case of positrons, they can annihilate with an electron), depositing their energy into the ejecta, heating it.  This energy can then power the light-curve.  For thermonuclear supernovae, $^{56}$Ni decay is the primary source of the observed thermal emission.

For core-collapse supernovae, the role of $^{56}$Ni is less clear.  For core-collapse explosions, the $^{56}$Ni is produced in the central engine and the energy must diffuse out to the photosphere to power the light-curve.  Mixing this $^{56}$Ni into the outer ejecta, the diffusion time can be reduced.  Unless there is extensive mixing (or the ejecta mass is low, less than roughly 1\,M$_\odot$), radioactive decay primarily powers the late-time light-curve~\cite{doi:10.1142/S0218271825400024}.  Mixing, coupled with the low-mass ejecta in type Ib/c SNe, the decay of $^{56}$Ni could power the peak of the light-curve although, unless the mass is low, it is unlikely to explain the fast decay of the SN emission~\cite{2025ApJ...994..259N}.

For kilonovae, transients formed through neutron star merger, a broad range of radioactive isotopes are produced whose decay can power the light-curve.  This decay energy is typically included through formulae that integrate over a broad range of radioactive isotopes~\cite{2016ApJ...829..110B,2021ApJ...918...44B}.  These radioactive isotopes are believed to be fully mixed across the ejecta.  Ultimately, accurate models of kilonova light-curves will require the full distribution of radioactive isotopes.

As we shall discuss in Sections~\ref{sec:g-ray} and \ref{sec:SNR}, we can directly observe the yields of these radioactive isotopes through the $\gamma$-rays emitted in their decay that escape without losing too much energy through scattering.  


\subsection{Thermal Shocks}

For core-collapse supernovae, shock heating can be a critical power source.  As the supernova blastwave propagates through the star and the surrounding circumstellar medium, the shock will decelerate and its kinetic energy will be converted into thermal energy in the shock.  This energy powers shock breakout, type II supernova light-curves,and a subset of Type Ib/c (including superluminous supernovae).  {\bf The importance of shock heating in supernovae remains a matter of debate.}

\begin{figure}
    \centering
    \includegraphics[width=5.5in]{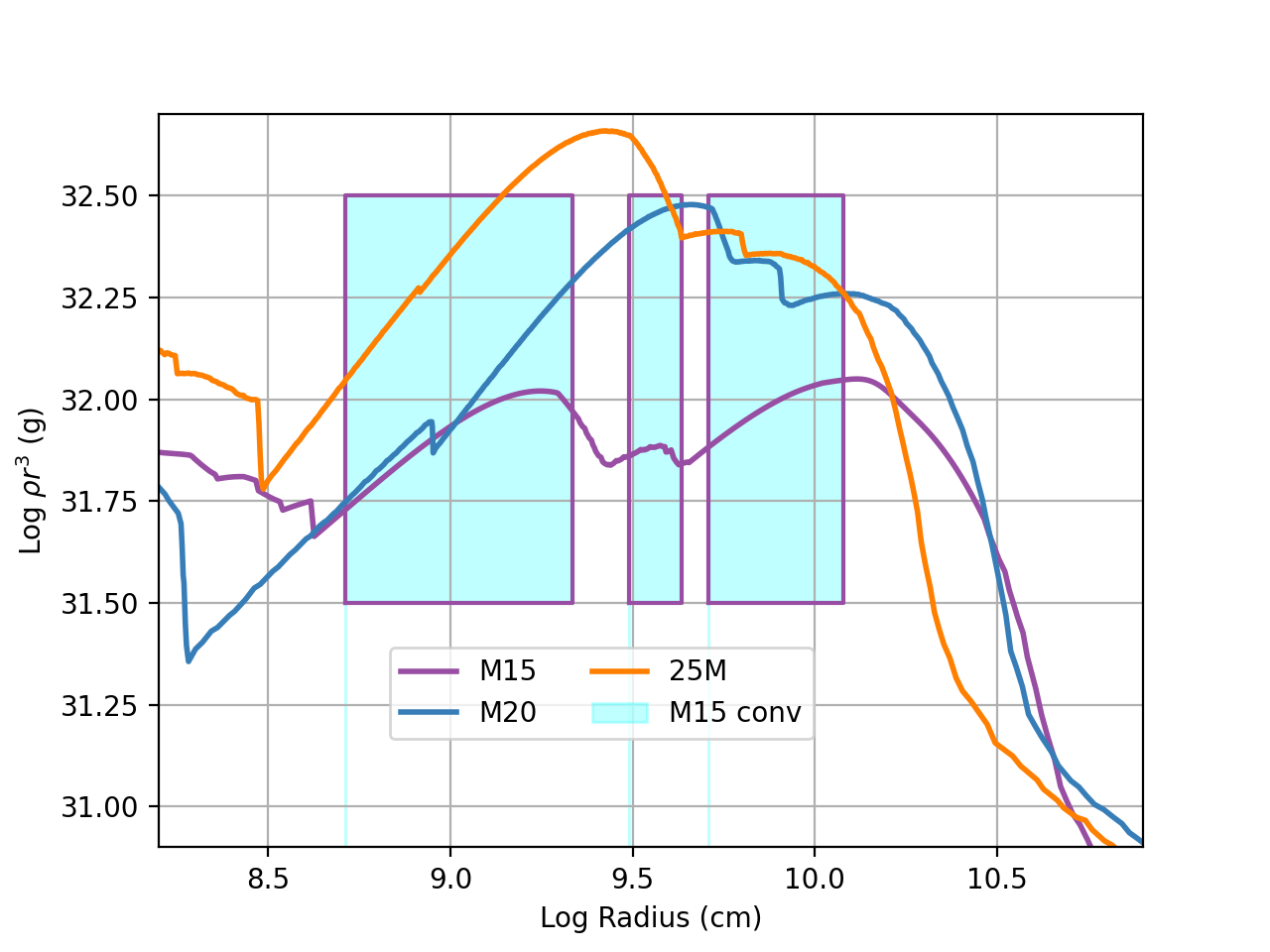}
    \caption{Density times radius cubed ($\rho r^3$) versus radius for 3 collapse progenitors, varying the stellar mass: 15, 20, 25\,M$_{\odot}$.  If this value increases with radius, $\alpha_\rho<3$ and the shock decelerates, causing kinetic energy to be converted to thermal energy.  If the value decreases, the shock will accelerate.  Note that $\rho r^3$ decreases dramatically at the edge of the star.  This will lead to rapid acceleration, possibly producing relativistic ejecta.  This is Figure 2 from \cite{2026arXiv260104464F}.}
    \label{fig:sedovstar}
\end{figure}

As the supernova blastwave moves out through the star, its velocity can both increase or decrease based on the density profile of the star.  This propagation is well-described by the Taylor–von Neumann–Sedov similarity solution~\cite{1959sdmm.book.....S}:
\begin{equation}
 v_{\rm shock} \propto v_0 (t/t_0)^{(\alpha_\rho-3)/(5-\alpha_\rho)}
\end{equation}
where $\alpha_\rho$ denotes the density gradient ($\rho \propto r^{-\alpha}$), $v_{0}$ is the initial velocity when the shock starts decelerating, and $t_0$ is the time at the onset of the deceleration.  If $\alpha>3$, the shock accelerates and if $\alpha<3$, it decelerates.  When the blast wave propagates through a star, it will pass through regions of the star where $\alpha$ is both greater and lower than 3 (see Figure~\ref{fig:sedovstar}).

This shock heating can be understood through a simple energetics solution.  The change of energy ($\Delta e$) density from the deceleration is:
\begin{equation}
    \Delta e = 1/2 \rho (v_{orig}^2 - v_{dec}^2) \, {\rm erg \, cm^{-3}}
\end{equation}
where $\rho$ is the density, $v_{orig}$ is the original velocity, and $v_{dec}$ is the velocity after the deceleration.  For astrophysical transients, the thermal energy is radiation dominated and the corresponding temperature at the shock is given by: 
\begin{equation}
    a (T_{\rm shock}^4 - T_{\rm orig}^4) \approx a T_{\rm shock}^4 = \Delta e
\end{equation}

Through much of the star, particularly in the stellar envelope, the blastwave will decelerate.  The energy released in this propagation heats the shock leading to temperatures of nearly 1 million Kelvin when the blastwave expands out of the star.  Material piles up as the shock decelerates.  This pile-up causes a so-called reverse shock to propagate into the ejecta (this reverse motion refers to an inward motion in mass, not radial, coordinate).  For giant stars, this shock-heated material will dominate the light-curve.  As the photosphere moves inward in mass, it uncovers material heated by the pile-up or reverse shock, ensuring a emission over long timescales.

Shock heating is also important in shock breakout (see Section~\ref{sec:SBO}).  The steep decrease at the edge of the star causes a thin layer of the blastwave to accelerate, ultimately reaching relativistic velocities.  When it decelerates as it hits the circumstellar medium, this shock-heated material can become so hot that its Lorentz-boosted emission can produce gamma-ray bursts~\cite{1968CaJPh..46..476C,2001ApJ...551..946T}.  

The shocks produced as the supernova blastwave propagates through the circumstellar medium can be a major power source for type Ib/c and superluminous supernovae~\cite{2017ApJ...850..133D,2017hsn..book..403S,2019ApJ...874...68C,2020ApJ...898..123F,2025ApJ...991...22F}.  We will discuss these sources in more detail in Section~\ref{sec:LC}

\subsection{Synchrotron Emission}

Shocks do not naturally produce a thermal distribution.  The assumption in our thermal shocks is that the ions, electrons and radiation all equilibrate, a.k.a. local thermodynamic equilibrium (LTE).  While this is a good approximation for the emission at the modest densities occurring near the peak of the supernova EM emission,  when the shock reaches low densities, fermi acceleration across the shock can produce a distribution of non-thermal electrons.    These high energy particles emit synchrotron emission as these high energy particles are accelerated in the surrounding magnetic fields.  

Synchrotron emission has been used to explain a wide set of transient observations:  gamma-rays from gamma-ray bursts as the jet propagates through the circumstellar medium roughly 0.1\,pc from the central engine~\cite{1998ApJ...497L..17S}, X-ray and radio emission in supernova remnants~\cite{2008ARA&A..46...89R}, emission from magnetized neutron stars~\cite{2002ApJ...574..332T}, high energy particles (neutrinos, cosmic rays) in all transient shocks.  This mechanism can only occur at low densities where thermalization timescales are long.  {\bf But beyond the non-thermal requirement, this mechanism can work in a wide range of conditions.}

Typically, astronomers assume that the fermi acceleration in these shocks produce a power-law distribution for the high energy electrons.  {\bf Observations require fairly flat indices for this power law ($-2--3$), something not generally achieved by plasma physics calculations, e.g. kinetic or particle-in-cell methods~\cite{2023ApJ...952..165R} although models can be produced~\cite{2023ApJ...952..165R,2026ApJ..1005..107F,2026ApJ...998..149J} (but note the assumptions for the magnetic field structure and electron to ion mass in these models).}  Much more work is required to gain a first-principles understanding of this energy source for astrophysical transients.  {\bf It is also important to bear in mind that alternate sources of these observations exist and the importance of this emission source remains a matter of debate.}

\subsection{Late-Time Central Engine Activity}

A final energy source lies in late-time energy extraction from the compact object itself.  This can occur either through the emergence of a magnetar that taps the rotational energy of a neutron star or it can be driven by further accretion onto the neutron star (a.k.a. ``fallback'').  Both of these power sources deposit energy into the innermost ejecta which must the diffuse out to the photosphere to produce the observed emission~\ref{fig:emsources}.  Let's study these two mechanisms in more detail.  {\bf These mechanisms have been proposed to explain exotic supernovae such as superluminous supernovae or fast blue optical transients.  Because alternative explanations of these exotic outbursts exist and because their relative rates are low, these mechanisms may not occur often.  For example, in many studies, the rate of superluminous supernovae is 0.01-0.1\% of the total supernovae rate~\cite{2021MNRAS.500.5142F} (note that this number remains a matter of debate).}

\subsubsection{Fallback}

As the supernova blastwave moves out through the star, a portion of its kinetic energy is used to eject the stellar material.  The loss of this energy can cause it to decelerate below the escape velocity of the compact remnant, causing it to fall back on the compact remnant~\cite{1971ApJ...163..221C,1989ApJ...346..847C,2009ApJ...699..409F}.  As this material falls back onto the neutron star, it develops an atmosphere on top of the neutron star.  This atmosphere is violently convective and this convection can drive outflows~\cite{1996ApJ...460..801F,2019MNRAS.485..620K,2026ApJ...997...88A}.  Before we discuss these multi-dimensional effects, lets understand the simple spherically-symmetric picture.

The entropy ($S$) of this atmosphere is set by the strength of the shock as it piles up on the neutron star.  The infall velocity (essentially the free-fall velocity) decreases as the atmosphere builds outward.  It also depends upon the accretion rate ($\dot{M}_{\rm fallback}$) by~\cite{1996ApJ...460..801F}:
\begin{equation}
    S = 1200 \dot{M}^{-1/4}_{\rm fallback} (M_{\odot}/s) r_6^{-3/8} k_B/{\rm nucleon} 
\end{equation}
for a 1.4\,M$_\odot$ neutron star where $r_6$ is the radius of this atmosphere.  The radial dependence on the entropy means that the atmosphere is initially unstable to Rayleigh-Taylor convection.  The growth rate of convection in this atmosphere can be approximated through the Brunt-V\"ais\"ala frequency~\cite{1983apum.conf.....C} that, in the radiation-pressure dominated conditions of our fallback atmospheres, can be approximated by~\cite{2007ApJ...659.1438F}:
\begin{equation}
    \omega^2 = \frac{G M_{\rm enclosed}}{r^2} \frac{1}{S} (\partial S/\partial r)
\end{equation}
where $G$ is the gravitational constant, $M_{\rm enclosed}$ is the enclosed mass (close to the neutron star mass) and $r$ is the radial position ($G M_{\rm enclosed}/r^2$ is the gravitational acceleration).  If $\omega^2$ is negative ($\partial S/\partial r < 0$), the atmosphere is unstable to Rayleigh-Taylor convection.  The growth of this convection is $\tau_{\rm RT}$ is then:
\begin{equation}
    \tau_{\rm RT} \approx |(1/\omega^2)^{1/2}| \approx 7 \times 10^{-5} (M_{\rm enclosed}/M_{\odot})^{-0.5} r_6^{1.5} s.
\end{equation}
Even out to 100\,km, the convective growth time is less than a millisecond~\cite{1996ApJ...460..801F}, on par with that expected in core-collapse models~\cite{2007ApJ...659.1438F}.  It is important to note that the numerical (artificial) viscosities in current simulations typically lead to growth times that are $>100$ times larger than our more-accurate analytic expectations.  This will cause the accreted material to quickly evolve into a constant-entropy, pressure equilibrium atmosphere.  The density ($\rho_{\rm atm}$) and temperature ($T_{\rm atm}$) profile on this 1.4\,M$_\odot$ neutron star is given by~\cite{1993PhR...227..157C}:
\begin{equation}
    \rho_{\rm atm} = 3.9 \times 10^{15} S^{-4} r_6^{-3} {\rm g \, cm^{-3}},
    \label{eq:rhoatm}
\end{equation}
\begin{equation}
    T_{\rm atm} = 195 S^{-1} r_6^{-1} {\rm MeV}.
    \label{eq:tatm}
\end{equation}
At the high temperatures and densities of this atmosphere, neutrino cooling (both pair-annihilation and, at higher densities, electron capture) decrease the entropy/energy in and, ultimately deleptonize this atmosphere, causing it to accrete onto the neutron star.  Neutrino cooling, predominantly from electron ($d\epsilon_{nuc}/dt$) or pair annihilation ($d\epsilon_{pp}/dt$), can be approximated by~\cite{1992ApJ...395..642H}:
\begin{equation}
    d\epsilon_{nuc}/dt \approx 2 \times 10^{18} T_{\rm MeV}^6 {\rm \, erg \, g^{-1} \, s^{-1}}
    \label{eq:denuc}
\end{equation}
and
\begin{equation}
    d\epsilon_{pp}/dt \approx 1.9 \times 10^{25} T_{\rm MeV}^9/\rho {\rm \; erg \, g^{-1} \, s^{-1}}
    \label{eq:depp}
\end{equation}
where $T_{\rm MeV}$ and $\rho_{\rm atm}$ are the temperature (in MeV), density (in ${\rm g \, cm^{-3}}$) of the atmosphere.  Combining equations \ref{eq:rhoatm}-\ref{eq:depp} in the integral, the net neutrino cooling is:
\begin{equation}
    L_{\nu} = \int_{r_{\rm NS}}^{r_{\rm atm}} 4 \pi r^2 \rho_{\rm atm} (d\epsilon_{nuc}/dt+d\epsilon_{pp}/dt) = S_{10}^{-10}(7.7\times10^{55}+1.4\times10^{55}S_{10})
\end{equation}
where $S_{10}$ is the entropy of the atmosphere in units of $10 \, k_{\rm B} {\rm \, per \, nucleon}$ where $k_{\rm B}$ is the Boltzmann constant.  The corresponding cooling time for neutrinos is then:
\begin{equation}
    t_{\nu} = \dot{E}_{\rm acc}/L_{\nu} = \dot{M}_{\rm acc} G M_{\rm NS}/r^2_{\rm NS} / L_{\nu}
\end{equation}

\begin{figure}
    \centering
    \includegraphics[width=5.5in]{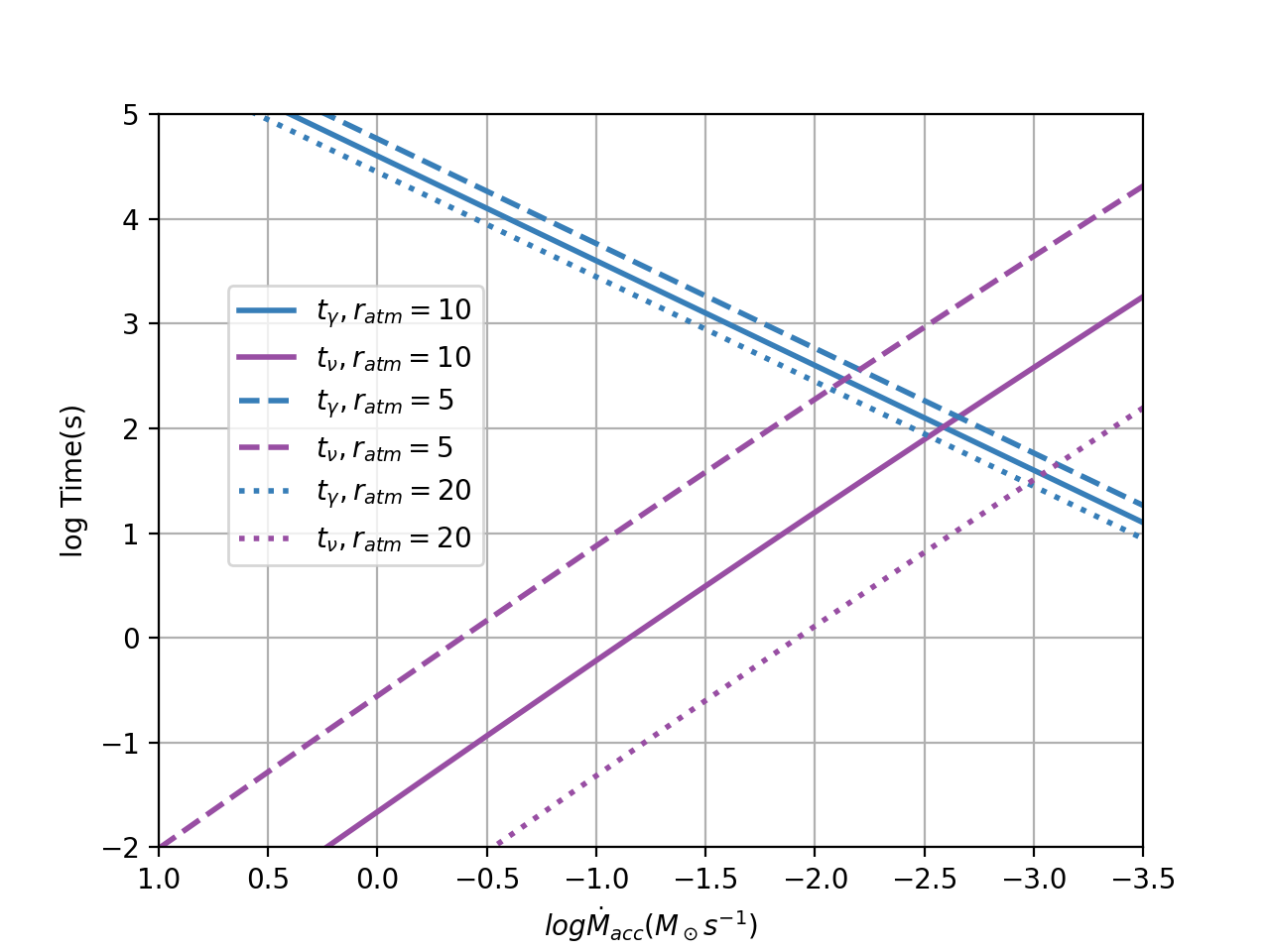}
    \caption{Neutrino cooling versus photon diffusion timescales for a range of accretion flows during supernova fallback for 3 different atmosphere sizes:  50, 100, and 200\,km.  The neutrino cooling timescale is the time required for neutrinos to release the gravitational potential released in the infall.  The diffusion timescale is the time required for energy at the neutron star surface to rise to the edge of the atmosphere.}
    \label{fig:ftime}
\end{figure}

If we focus on a stable, spherically-symmetric envelope, photons can only alter this accretion (and add power to the observed EM emission) if they can diffuse out of this envelope before neutrinos cool the atmosphere.  The optical depth of the atmosphere is given by:
\begin{equation}
    \tau = \int_{r_{\rm PNS}}^{r_{\rm atm}} \sigma \rho_{\rm atm}(r) dr
\end{equation}
where $\sigma \approx 0.2 {\rm \, cm^2 \, g^{-1}}$ is the opacity of the atmosphere (electron scattering for a $Y_e\approx 0.5$ composition).   The diffusion time through this atmosphere is more difficult to estimate.  Here we assume an average mean free path is $\lambda_{\rm ave} = \Delta r/\tau$.  Then the timescale ($t_{\gamma}$) is roughly:
\begin{equation}
    t_{\gamma} \approx \tau^2 \lambda_{ave}/c = \tau \Delta r/c
\end{equation}
Figure~\ref{fig:ftime} shows the diffusion timescale versus the neutrino cooling timescale as a function of the accretion rate.  Neutrino cooling is fast and the density is so high that the photons are trapped in the flow.  But as the accretion rate decreases, the density drops sufficiently that the photon diffusion timescale is shorter than neutrino cooling.   

Initially, the accretion rate is high (0.1-1\,M$_{\odot} \,  s^{-1}$), but it decreases with time.  At early times, the fallback accretion rate can be chaotic~\cite{2009ApJ...699..409F,2025MNRAS.538..572S} (indeed, with asymmetric explosions, material keeps on falling onto the neutron star even when the explosion is being launched).  At late times, the accretion rate will become more smooth, typically follows the $t^{-5/3}$~\cite{1989ApJ...346..847C}, but asymmetric explosions are likely to produce more complex accretion structures.  As such, the exact accretion rate where photons effectively diffuse out to power late-time light-curves is difficult to determine exactly.  

\begin{figure}
    \centering
    \includegraphics[width=5.5in]{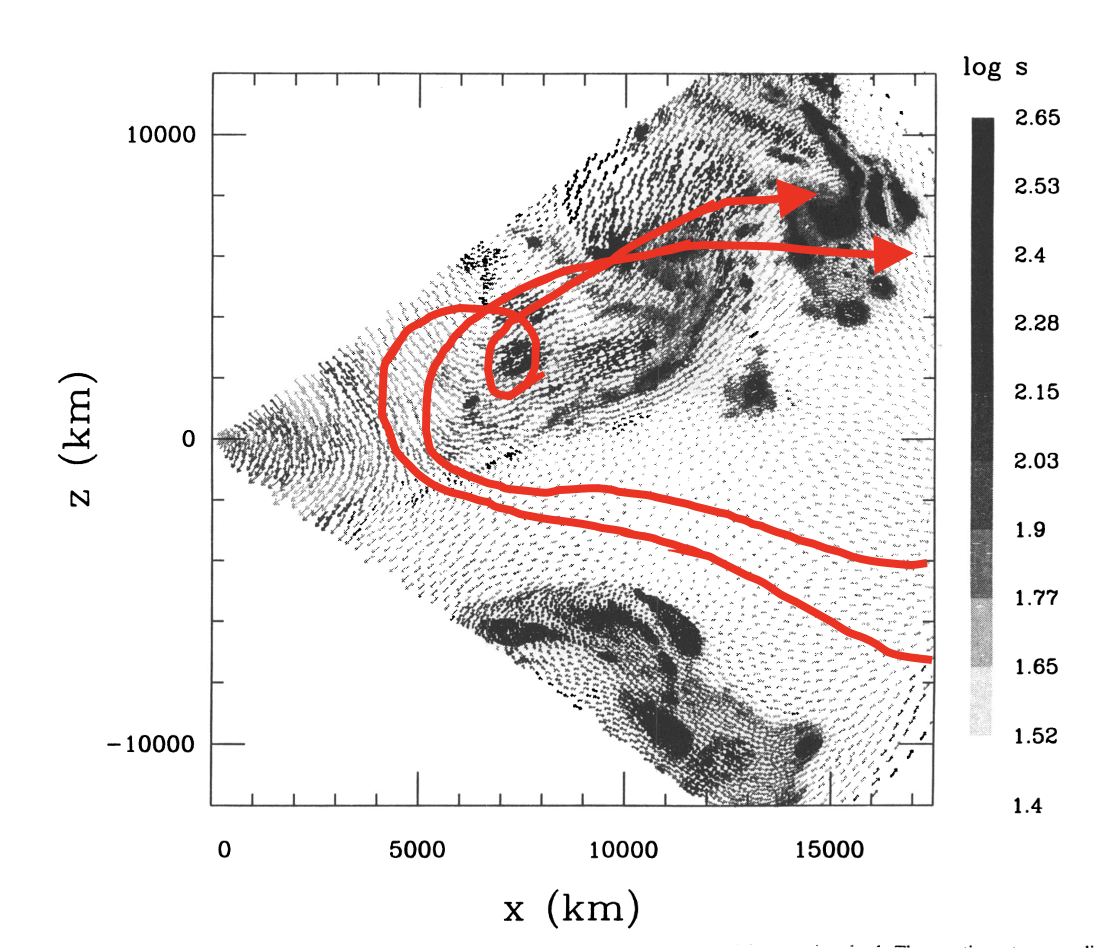}
    \caption{Sample inflow trajectories plotted on top of a single time snapshot of fallback and outflow models~\cite{1994ApJ...435..339H}.  The nature of these trajectories affects both the electron fraction and the time duration at high temperatures.  Both electron fraction and trajectory differences are behind the different yields from ~\cite{2019MNRAS.485..620K} and ~\cite{2026ApJ...997...88A}.}
    \label{fig:FBeje}
\end{figure}

Although much of the material falling down onto the neutron star will accrete, some can be reheated and ejected (Figure~\ref{fig:FBeje}).  This ejected material can be processed through extreme conditions, producing heavy elements either through rapid proton- or neutron-processes~\cite{2006ApJ...646L.131F,2019MNRAS.485..620K,2026ApJ...997...88A}.  The nature of this nucleosynthesis depends on the nature of the flow shown in Figure~\ref{fig:FBeje} and how long the inflowing material spends in extreme conditions.  For instance, ~\cite{2019MNRAS.485..620K} assumed a simple inflow/outflow trajectory that spent little time at extreme conditions.  ~\cite{2006ApJ...646L.131F,2026ApJ...997...88A} had flows that would cycle, spending more time at high temperatures.  The high-temperatures/densities not only allowed more nuclear burning, but could also lead to neutrino emission and deleptonization, leading to r-process nucleosynthesis~\cite{2006ApJ...646L.131F,2026ApJ...997...88A}.  Deleptonization is also sensitive to the nature of the neutrino microphysics and detailed numerics for the neutrino transport.  To obtain accurate yields and energetics from these outflows, much more work is needed.  But, in all cases, these flows will increase the amount of energy deposited into the explosion to power light-curves at late-times and observing the late-time emission can constrain this physics.

\subsubsection{Magnetar Energies}

Another central engine source invokes tapping the rotational energy of a neutron star remnant.  Magnetar engines tap the rotational energy in the newly-formed NSs to power emission or a jet.  Analytic estimates can provide some insight into the importance of this energy-injection mechanism.

Fundamental physics already places limits on this energy.  The moment of inertia for neutron stars depends upon the equation of state, but all estimates of the moment for a NS ($I_{\rm NS}$) are within a factor of two of~\cite{2008ApJ...685..390W}:
\begin{equation}
I_{\rm NS} = 10^{45} (M_{\rm NS}/M_\odot) {\rm~g~cm^2}
\end{equation}
where $M_{\rm NS}$ is the NS mass. The corresponding
rotational energy ($E_{\rm rot}$) is:
\begin{equation}
    E_{\rm rot} = 5 \times 10^{50} (\omega/1000\,{\rm Hz})^2 \rm{erg}
\end{equation}
where $\omega$ is the angular velocity.  A NS with a spin period of 1\,ms has a total of $2 \times 10^{52} {\rm \, erg}$.  

The maximum spin period of NSs is limited by instabilities in the NS.  The onset of these instabilities occurs when the rotational energy exceeds 14\% of the potential energy of the neutron star~\cite{1983bhwd.book.....S}:
\begin{equation}
\beta = E_{\rm rot}/|W| > 0.14
\end{equation}
where 
\begin{equation}
    |W| = G M_{\rm NS}^2/r_{\rm NS} \approx 5 \times 10^{53}\,{\rm erg}
\end{equation}
and, for the neutron star mass and radius, we have $M_{\rm NS}=1.4\,M_\odot, r_{\rm NS} = 10\,{\rm km}$, respectively.  These instabilities place an upper limit on the energy available for a magnetar of $\approx 7 \times 10^{52} {\rm erg}$.  But other instabilities can further reduce the maximum total energy.  For example, at a fixed angular momentum, the rotational energy in the proto-neutron star increases as it becomes more compact.  If the magnetar-strength magnetic fields develop when the proto-neutron star is still hot and extended, it can lose its angular momentum before the rotational energy reaches its peak.  For a given amount of total angular momentum ($J_{\rm tot}$), the total rotational energy is inversely proportional to the square of the radius:
\begin{equation}
    E_{\rm rot} = 1/2 I_{\rm NS} \omega^2 = 1/2 J_{\rm tot}^2/I_{\rm NS} \propto J_{\rm tot}^2/r_{\rm NS}^2.
\end{equation}
A hot NS has a radius roughly 3 times that of a cold NS.  At this point, the total rotational energy is roughly 10 times lower than it will have when it cools.  For BH forming systems, this is the only magnetar energy reservoir, making it difficult for magnetar engines to have much power for these systems.  The most powerful magnetars are likely to be produced in NS-forming systems where the magnetar-like fields are not completely formed until the NS cools.

\begin{figure}
    \includegraphics[width=0.9\textwidth]{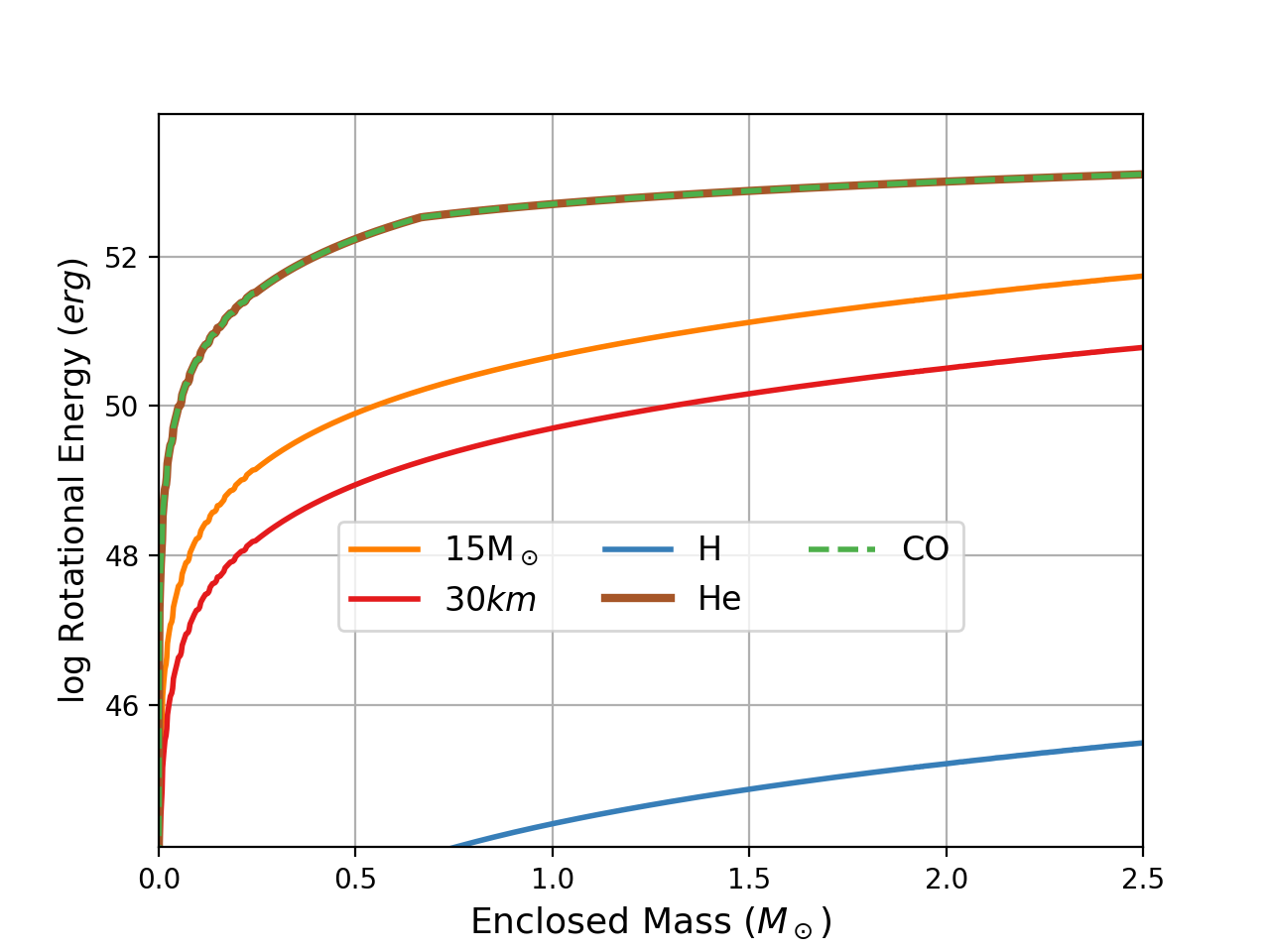}
    \caption{Available rotational energy for neutron stars in stellar collapse as a function of zero-age main sequence mass for both single stars and binary systems.  For the single system, we consider a weakly coupled star where the magnetar forms after the neutron star has cooled or when it is still hot with a radius of $30\, {\rm km}$.  For our binary systems, we asume a tight binary where the star is tidally spun up as a H, He or CO star where the star is very close to the Roche radius.} 
    \label{fig:mage}
\end{figure}

Figure~\ref{fig:mage} shows the range of energies for both single stars and binary systems.  The angular momentum in a star is dictated by both angular momentum coupling between the different composition layers in the star and spin-up in binaries.  A wide range of angular momentum coupling prescriptions are used in stellar evolution (for a review, see~\cite{2020A&A...636A.104B}) producing a wide range of pre-collapse angular-momentum profiles.  We include models with both a weak and strong coupling.  For weak coupling assuming Tayler-Spruit dynamo~\cite{2002A&A...381..923S}, the available rotational energy can be $10^{51}\,{\rm erg}$ for a 1.4\,M$_\odot$ neutron star.  If the magnetar forms early before the neutron star has cooled, the total energy is closer to $10^{50}\,{\rm erg}$.  If the coupling is strong, the total energy is less than $10^{46}\,{\rm erg}$.  However, if the star is in a tight binary system, tidal forces can spin up the star.  In this case, the collapsed neutron star can have considerable rotational energy that exceeds $10^{52}\,{\rm erg}$.

These extreme energies can be over-estimates.  If Rossby waves or similar instabilities develop neutron star, the rotational energy would be lost through gravitational waves~\cite{2003ApJ...591.1129A,2006ApJ...651.1068O,2010ApJ...709...77M,2011LRR....14....1F} and the remaining energy available to magnetars would be low.  But viscous damping of these waves can limit the amount of angular momentum lost to these waves~\cite{2000ApJ...529L..33B}.  If we ignore Rossby instabilities, focusing only on other (e.g. bar) instabilities, we can estimate the maximum rotational energy available for magnetar engines.

From these results, we can identify trends in SNe from magnetars.  For our single star models assuming coupling caused by the Tayler-Spruit dynamo~\cite{2002A&A...381..923S}, the magnetar energies are typically below $10^{50}-10^{51} {\rm \, erg}$.  Observations suggest that, for most systems, the spins are even lower (see Section~\ref{sec:compactremnants}).  The most powerful magnetars are likely to arise from rare systems, e.g. tightly-coupled binaries.  A broad range of CO binaries achieve rotational energies near the maximum value derived above.  Under these progenitors, we would expect most magnetar-driven SNe to be classified as type Ib/c SNe.  But even if the magnetar isn't the primary drive of the supernova, magnetars could provide additional energy to the explosion, powering late-time light-curve emission.  

\section{Observations}
\label{sec:observations}

Although gravitational waves and neutrinos are the most direct probe of the supernova engine, they are rare.  Even the most rare EM observations are likely to grow more quickly than these direct probes over the next few decades.  A broad range of EM observations exist spanning from observations of the initial explosion (shock breakout) to the study the explosive remnants (both compact and ejecta).  These observations all pose their own challenges.  Table~\ref{tab:observations} summarizes these different observations, the aspects of the supernova engine and its progenitor that they constrain, and the challenges in using these progenitors.

\begin{table}
\caption{Description}
\centering
\begin{tabular}{| l | l | l | }
\hline
Diagnostic & Constraint & Challenges \\
\hline
Shock Breakout & Stellar Radius, Circumstellar Medium & Solution Degeneracy \\
 & Explosion Properties & \\
Transient Light Curve & Ejecta Mass, Velocity (magnitude and & Difficult to Disentangle \\
& distribution), Power Source & Properties \\
Transient Spectra & Composition, Velocity & Photosphere and NLTE effects \\
Transient $\gamma-$rays & Conditions in Engine & Limited Number of Systems \\
 & (energies, asymmetries) & with Sufficient Signal \\
Ejecta Remnants & Composition, Energy & Must Disentangle from  \\
& & Circumstellar Medium Effects \\
Compact Remnants & Angular Momentum, Explosion & Need Theory Models \\
\hline
\end{tabular}
\label{tab:observations}
\end{table}

\begin{figure}
    \includegraphics[width=0.9\textwidth]{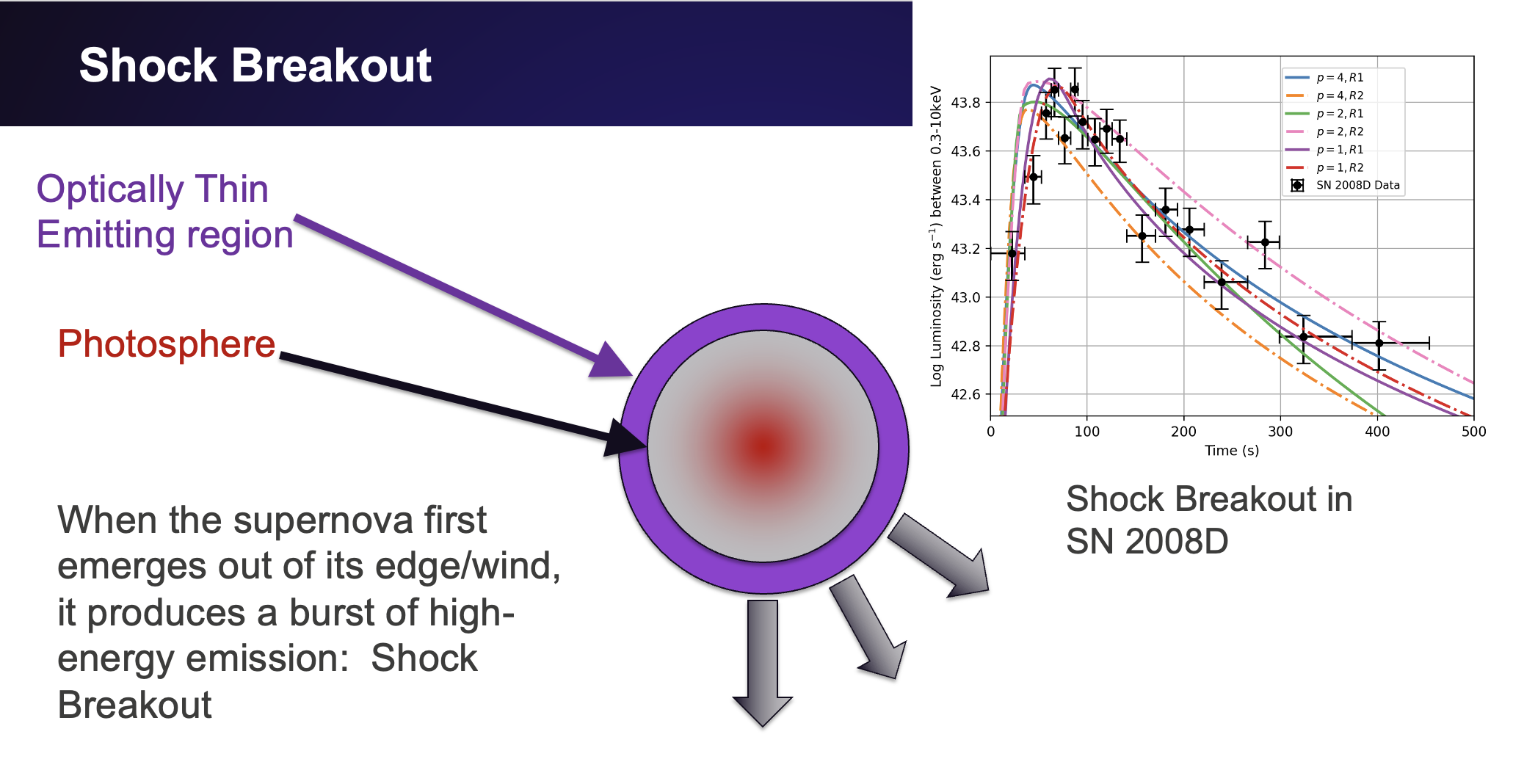}
    \caption{Conditions at shock breakout.  The shock emerges from the star or stellar wind and the ejecta is just getting sufficiently diffuse such that radiation can lead the material shock.  Shock heating dominates the observables.} 
    \label{fig:sbopic}
\end{figure}

\begin{figure}
    \includegraphics[width=0.9\textwidth]{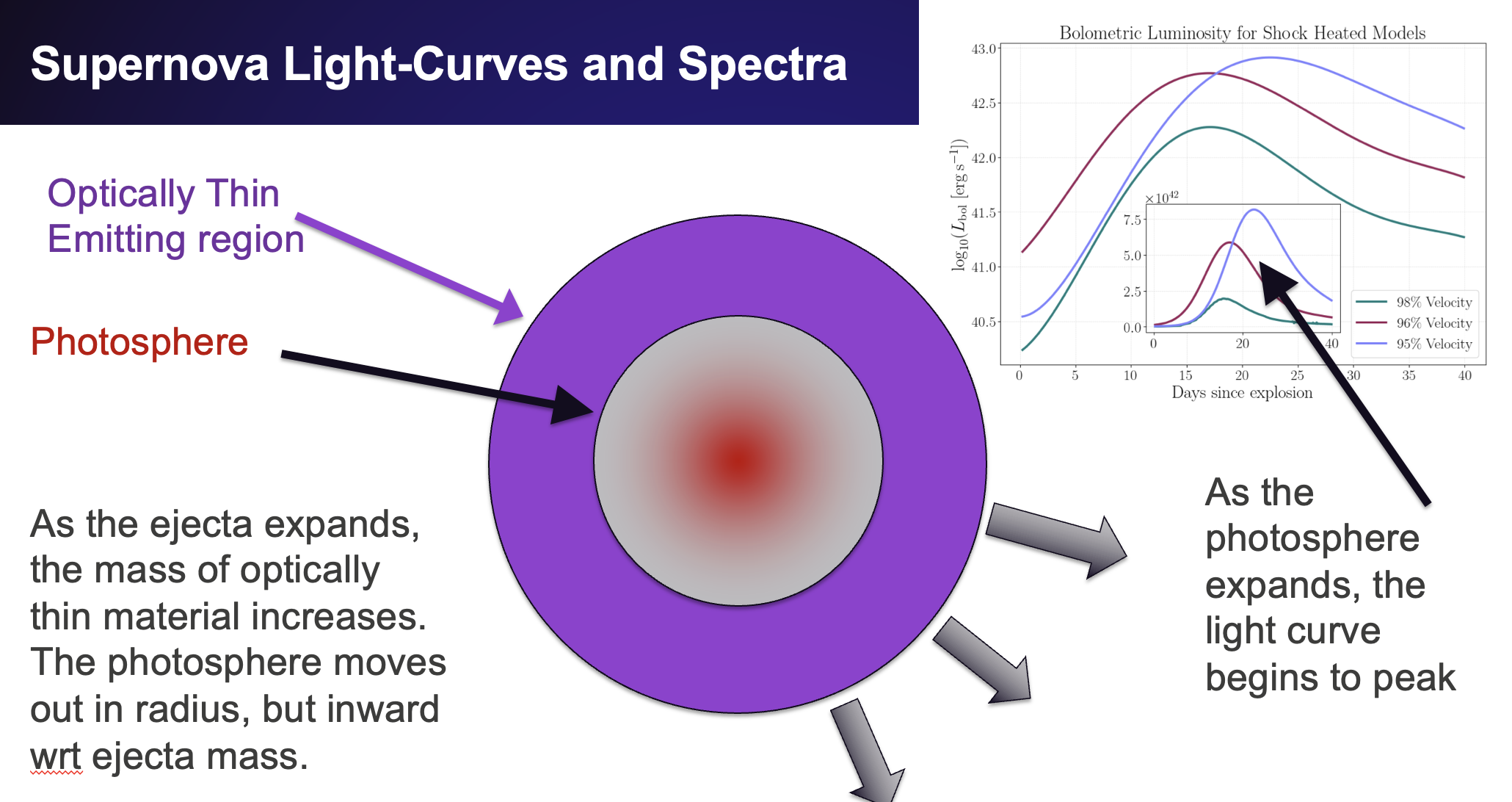}
    \caption{Conditions near the peak of the supernova emission.  Emission from the photosphere dominates the observations coupled with line features from the material above the photosphere.  The radiation from interior sources (radioactive decay, central engine) must diffuse out of the star to contribute to the light-curve.  } 
    \label{fig:SNLC}
\end{figure}

\begin{figure}
    \includegraphics[width=0.9\textwidth]{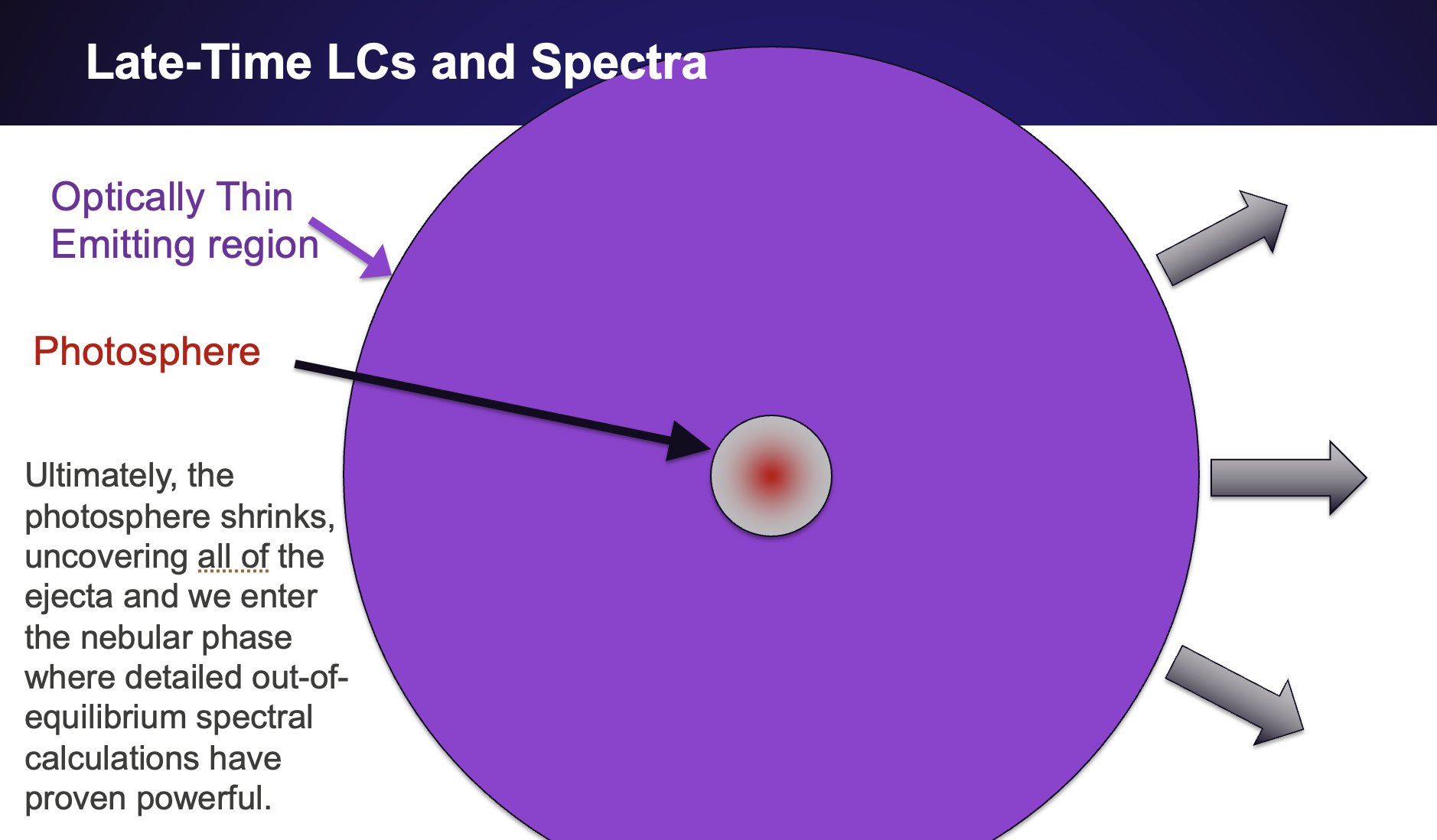}
    \caption{At late times, most of the ejecta is beyond the photosphere.  Spectra can probe the entire ejecta composition, but the conditions are far from equilibrium.} 
    \label{fig:latetime}
\end{figure}

We can divide the EM constraints into two classes:  transient observations and remnant observations.  Transient observations are differentiated both in time and photon energy.  The first emission, primarily in the UV and X-rays, is shock breakout (Figure~\ref{fig:sbopic}). As the blastwave expands, the emission peaks across UVOIR bands (Figure~\ref{fig:SNLC}).  Time-dependent band emission (light-curves) and spectra during this peak can be used to probe the explosion properties, but spectra observations at late times (Figure~\ref{fig:latetime}) have proven to be powerful diagnostics.  Prompt $\gamma-$rays from these transients provide a nearly direct probe of this engine.  Our first 4 subsections will cover these emission processes.

After the initial transient, the ejecta continues to propagate through the interstellar medium.  The shocks produce continue to cause the ejecta to emit primarily in the radio and X-ray.  The decay of radioactive isotopes (for SNe, $^{44}$Ti) produce $\gamma-$ray emission.  Compact remnants, probed by X-ray, radio and even gravitational waves, also constrain the nature of the supernova explosion.  Our final 2 subsections discuss these diagnostics.

\subsection{Shock Breakout}
\label{sec:SBO}

Shock breakout is the term used to describe the immediate burst of high energy (typically in the UV or X-ray) photons that occurs when the supernova blastwave breaks out of the star.  At breakout, the photons at the front of this ejecta are no longer trapped and, as they escape, the produce a flash of emission (Figure~\ref{fig:sbopic}).   The standard picture of shock breakout assumes that only the tip of the forward shock of radius ($\delta r$) is responsible for this emission~\cite{2017hsn..book..967W}. We can determine the onset of shock breakout when the radiative velocity in this ejecta (approximated by the diffusion velocity - $v_{\rm diff}$) exceeds the blastwave velocity ($v_{\rm bw}$).  The diffusive timescale is given by:
\begin{equation}
t_{\rm diff} = \lambda_{\rm mfp}/c (\delta r / \lambda_{\rm mfp})^2 = \delta r^2 \kappa \rho/c
\end{equation}
where $c$ is the speed of light and $\lambda_{\rm mfp} = 1/(\kappa \rho)$ is the mean free path where $\kappa$ is the opacity of the material (in ${\rm cm^2 \, g^{-1}}$) and $\rho$ is the density of the blastwave.  The diffusion velocity is then:
\begin{equation}
    v_{\rm diff} = \delta r/t_{\rm diff} = c/(\kappa \rho \delta r).
\end{equation}
By setting the radiation velocity to the blastwave velocity, we can determine the extent of $\delta r$:
\begin{equation}
    \delta r = (c/v_{\rm bw}) \kappa^{-1} \rho^{-1}
\end{equation}
and the mass enclosed ($M_{\rm bw}$) in this breakout region is roughly:
\begin{equation}
    M_{\rm bw} \sim \rho R_{\rm star}^2 \delta r = (c/v_{\rm bw}) \kappa^{-1} R_{\rm star}^2
\end{equation}
where $R_{\rm star}$ is the radius where shock breakout occurs, roughly the radius of the star.  In this simple model, we can use the emission timescale and peak flux to infer the radius of the star and the velocity of the shock.  {\bf This simple model predicts that, for a typical $10^{11} {\rm cm}$ Wolf-Rayet star and a blastwave velocity of $3 \times 10^9 {\rm cm \, s^{-1}}$, the diffusion timescale is $\sim 300{\rm s}$.}

But the real picture of shock breakout is much more complex than this simple model and recent studies are just starting to explore these complexities (\cite{2011AstL...37..194B,2013MNRAS.429.3181T,2017ApJ...850..133D,2020ApJ...898..123F}).  For example, we have assumed that the opacity is constant across the photon energy spectrum, but we know variations exist due to the density and temperature conditions (e.g. \cite{2015ApJ...805...98B}).  Including the wavelength dependence and out of equilibrium effects will alter these results~\cite{1978ApJ...223L.109K,2007ApJ...664.1026W,2010ApJ...719..881S,2013ApJS..204...16F}.  In this case, focusing only on a narrow region at the front of the shock does not capture the observed SBO signal because this emission is produced across a much broader region.

Another complexity is that the velocity of SBO is not limited to the peak velocity of the forward blastwave within the star.  For example, as the shock front slows down, it creates a reverse shock that moves through the inner ejecta.  This means that the shock heated region can be much more extended than a simple shock-front model assumes.  In addition, the stellar edge is poorly understood and its profile can dramatically affect the SBO signal~\cite{2015ApJ...805...98B,2017ApJ...845..103L}.  The stellar envelope and its edge can be inhomogeneous~\cite{2022ApJ...933..164G} causing the breakout to occur at different times at different parts of the star and drives a set of auxiliary shocks that further heat the shock.  Both of these effects extend the emission from SBO. Finally, explosion asymmetries from the engine also alter the SBO properties~\cite{2021MNRAS.508.5766I} and these asymmetries can also produce additional shocks.  All of these effects make it difficult to disentangle cooling flows from SBO effects.  As the extended cooling phase and true SBO are extremely difficult to disentangle in observations, we will include both of them in this paper. 

SBO effects can also be confused with other shock interactions that produce additional shock outbreak properties.  A complete study of SBO should include these broader set of emergent shock phenomena.  Examples of these studies include shock interactions in a clumpy wind medium (e.g. \cite{2020ApJ...898..123F}), interactions with mass-loss just prior to collapse (e.g. \cite{2017hsn..book..403S}), and shock interactions with companion stars (e.g. \cite{2010ApJ...717..245K}).  These shock interactions are probes of the circumstellar medium and not the stellar structure or explosion mechanism (for a review, see \cite{2025ApJ...994..259N}.  If these shocks occur at early times, they can be misinterpreted as SBO or contribute to the observed SBO signal. Table~\ref{tab:physicsmodels} lists the different SBO and broader shock interaction physics and their dominant effects on the SBO signal.

\begin{table*}
\centering
\begin{tabular}{|c|l|l|l|}
\hline\hline
 & \textbf{Process} & \textbf{Effects Beyond the Basic Model} &  \textbf{Model} \\
 &  &  &  \textbf{Implementation} \\
\hline
SBO & Forward Shock (basic model) & -- & \\
 Physics  & & & \\
& Shock Acceleration & Higher velocities and emission temperatures &  $E(\Gamma \beta)$ \\
 & Stellar Asymmetries & Expands shock-heated region and broadens rise time & $E(\Gamma \beta)$ \\
 & SN Asymmetries & Expands shock-heated region and broadens rise time & $A_{\rm emit}(\Gamma \beta,t)$  \\
 & Reverse Shock & Expands shock-heated region & $E_0$ \\
 & Radiation-driven Shocks & Produces a low-mass shock leading the material shock & Not Included \\
\hline
Shocks
 & & & \\
 & Clumpy Medium & Stochastic emission component & \\
 & & distributed heating broadening emission timescale & $\dot{E}_{\rm heat}$ \\
 & Shells & Prompt energy injection, sudden temperature rise & $\dot{E}_{\rm heat}$ \\
 & Companion Star & Prompt, localized energy injection & $\dot{E}_{\rm heat}$  \\
\hline\hline
\end{tabular}
\caption{Current analytical models only address the forward shock physics, ignoring many other processes that are known to occur.  Our implementation of the physics in our semi-analytical model includes a distribution of ejecta velocities, $E(\Gamma \beta)$, total energy reservoir, $E_0$, and emitting areas, $A_{\rm emit}(\Gamma \beta, t)$.  We also include a heating term, $\dot{E}_{\rm heat}$, to mimic late-time shock interactions. The details of our prescription are discussed in more detail in Section~\ref{sec:SBO}).}
\label{tab:physicsmodels}
\end{table*}

\begin{figure}
    \includegraphics[width=0.9\textwidth]{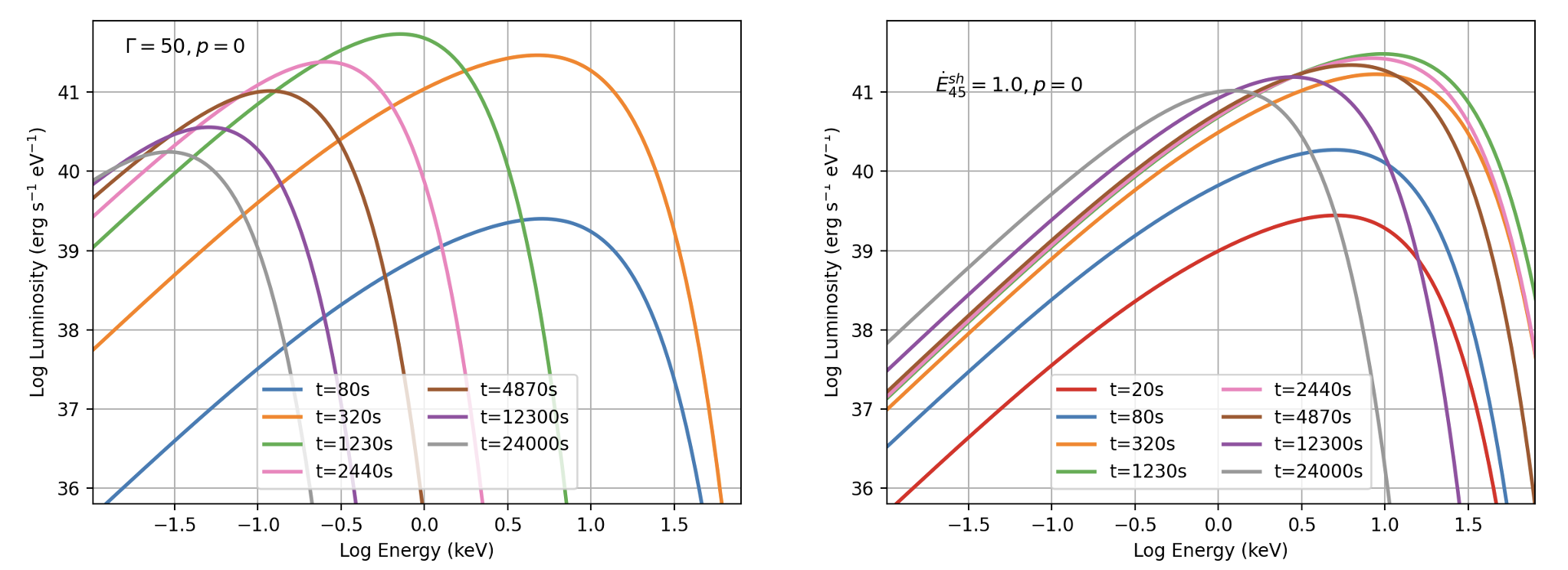}
    \caption{Spectra at a range of times with different Lorentz factors from shock breakout for 2 models from \cite{2026arXiv260300820F}.  The left image shows a mode that is more appropriate of a jet/disk model where the peak Lorentz factor reaches 50 (designed to match GRB 060218).  The right model has a lower peak Lorentz factor (10), but includes a modest amount of shock heating.   These are from Figure 19 of \cite{2026arXiv260300820F}.} 
    \label{fig:sbospectra}
\end{figure}

Figure~\ref{fig:sbospectra} shows time-dependent spectra from two different explosion models studied by~\cite{2026arXiv260300820F}.  For these energetic explosions, the emission peaks in the X-rays and then cools.  These spectra are expected to arise from Ib/c or broad-line Ic supernova explosions.   Such bright bursts could be observed by wide-field X-ray telescopes such as the Einstein Probe~\cite{2015arXiv150607735Y}.  Type II supernovae will peak in the UV and telescopes such as UltraSat~\cite{2024ApJ...964...74S} and UVEX~\cite{2021arXiv211115608K} will detect these SBO signals.

\subsection{Transient Light Curve}
\label{sec:LC}

By far, the most abundant data from supernovae comes in the form of supernova UVOIR light-curves:  emission (typically in broad energy bands) as a function of time.  Unfortunately, it is difficult to infer properties of the explosion from light curve models alone.  The simple model for supernovae light-curves assumes that the emission is powered by the decay of $^{56}$Ni and the subsequent decay of its daughter product $^{56}$Co.  Each releases energy in the form of $\gamma$-rays and positrons.  At early times, both are trapped in the ejecta and they quickly thermalize.  This thermalization energy heats the ejecta, causing it to radiate.  As this radiation reaches the photosphere (position where the radiation is no longer trapped), it escapes, producing a transient light curve (Figures~\ref{fig:SNLC},\ref{fig:latetime}).  Depending on the distribution of this $^{56}$Ni, one can derive the emergence of this energy (as it reaches the photosphere of the ejecta) as a function of time~\cite{1982ApJ...253..785A,2017ApJ...846...33A}.  

As with the shock breakout, this often-used model is far too simple to infer details of the explosion (ejecta mass, $^{56}$Ni mass, explosion energies).  The problem with the simple model is that it ignores many of the dependencies on the details of the explosion.  There is a misconception if a transient looks like a supernova (e.g. a 10-20 day rise time), it must be a $^{56}$Ni-driven supernova when, in fact, the light-curve says more about the mass and velocity of the ejecta~\cite{fryer224}.  First off, the decay of $^{56}$Ni is not the only way to power a supernova (see next paragraph).  Indeed, the ejecta of hot material, no matter how it is ejected, can produce a supernova-like outburst.  Supernova impostors arise from massive stars ejecting part of their envelope and these outbursts are often mistaken from supernovae (see, for example,  ~\cite{2009ApJ...697L..49S,2000PASP..112.1532V}). 

Supernova impostors are powered by ejecta made hot through internal shocks.   The difference between a supernova and a supernova impostor is not the energy source, but the cause of the ejection.  Core-collapse supernovae are produced in the collapse of the stellar core.  A number energy sources exist to power normal supernovae (Section~\ref{sec:sources}):  
\begin{itemize}
    \item{\bf Internal shocks:}  Type II supernovae can be powered primarily by internal shocks as the blast wave propagates through the star (pre shock breakout), similar to the power source of supernova impostors.  Accurate models of these light-curves primarily require capturing the photospheric radius and knowing the initial entropy distribution.
    \item {\bf Radioactive Decay}:  Type Ia supernovae and kilonovae are all likely to be powered primarily by the decay of radioactive isotopes.  For Type Ia supernovae, the primary energy is the decay of $^{56}$Ni to $^{56}$Co that in, turn, decays to $^{56}$Fe.  The gamma-rays and positrons released in this decay deposit energy into the ejecta.  For kilonovae, r-process elements decay emitting gamma-rays, electrons ($\beta$ particles) and $^4$He ($\alpha$ particles) which all deposit energy into the ejecta.  For kilonovae, the r-process elements are spread throught the ejecta, adding energy to all regions of the ejecta.  As the outermost regions become less dense, the gamma-rays (and then $\beta$ and $\alpha$ particles) are no longer trapped in the flow and the deposition of energy is reduced.  For thermonuclear supernovae, most of the $^{56}$Ni is produced in the inner half of the explosion.  Initially, its energy must diffuse out to power the light curve.  Type Ib/c supernova may also be powered by $^{56}$Ni decay, but the radioactive isotopes tend to be deeper in the ejecta and, because the ejecta is more massive, the decay energy doesn't reach the photosphere until late times.  Either this material must mix outward (even more so than it did in Supernova 1987A~\cite{1988Natur.333..534P,2005ApJ...635..487H}) or this power source will not drive the peak emission.  It will become more important at late times.  
    \item{\bf Shock Heating:}  Another power source for type Ib/c supernovae is shock heating as the supernova blast wave propagates through the dense stellar wind material of the Ib/c Wolf-Rayet progenitor.  As the blast wave moves out through this circumstellar material, it shock-heats and decelerates.  The deceleration can drive a reverse shock through the ejecta as it piles up on itself driving further shocks.  This shock heating becomes even more complex from asymmetric explosions that drive a broad set of secondary shocks through the ejecta all adding entropy to the explosion.  These shocks primarily occur in the outermost ejecta near the photosphere and directly deposit their energy into the ejecta that is also powering the observed emission.
    \item{\bf Magnetars or Continued Accretion:}  Many transients produce a central compact object (neutron star or black hole) and this formation event plays an important role in the engine and the launch of the initial transient blast wave.  But this central compact object can continue to add energy into the ejecta either through continued accretion (e.g. fallback~\cite{2014AIPA....4d1014F}) or tapping the rotation of a newly formed magnetar~\cite{2002ApJ...568..807W}.  This energy is deposited in the innermost ejecta and must diffuse out to affect the light-curves.  It is important in the late-time emission.  Only at very late times will we be able to distinguish magnetar from fallback energy sources and we will model them as a single entity in this paper.
\end{itemize}

In addition to the varying energy sources, a number of additional uncertainties can affect the light-curve.  For example, for $^{56}$Ni-powered supernovae, the distribution of the $^{56}$Ni can dramatically affect the light curve.  Although the $^{56}$Ni is produced in the innermost ejecta (created in or near the supernova engine), asymmetries in the explosion can cause it to be more broadly distributed in the ejecta.  Some light-curve models have implemented this by introducing a mixing parameter~\cite{2023ApJ...956...19F,2025ApJ...994..259N}.  Even the distribution of the ejecta mass (in velocity space) can dramatically alter the behavior of the light-curve~\cite{fryer224}.  

\begin{figure}
    \includegraphics[width=0.9\textwidth]{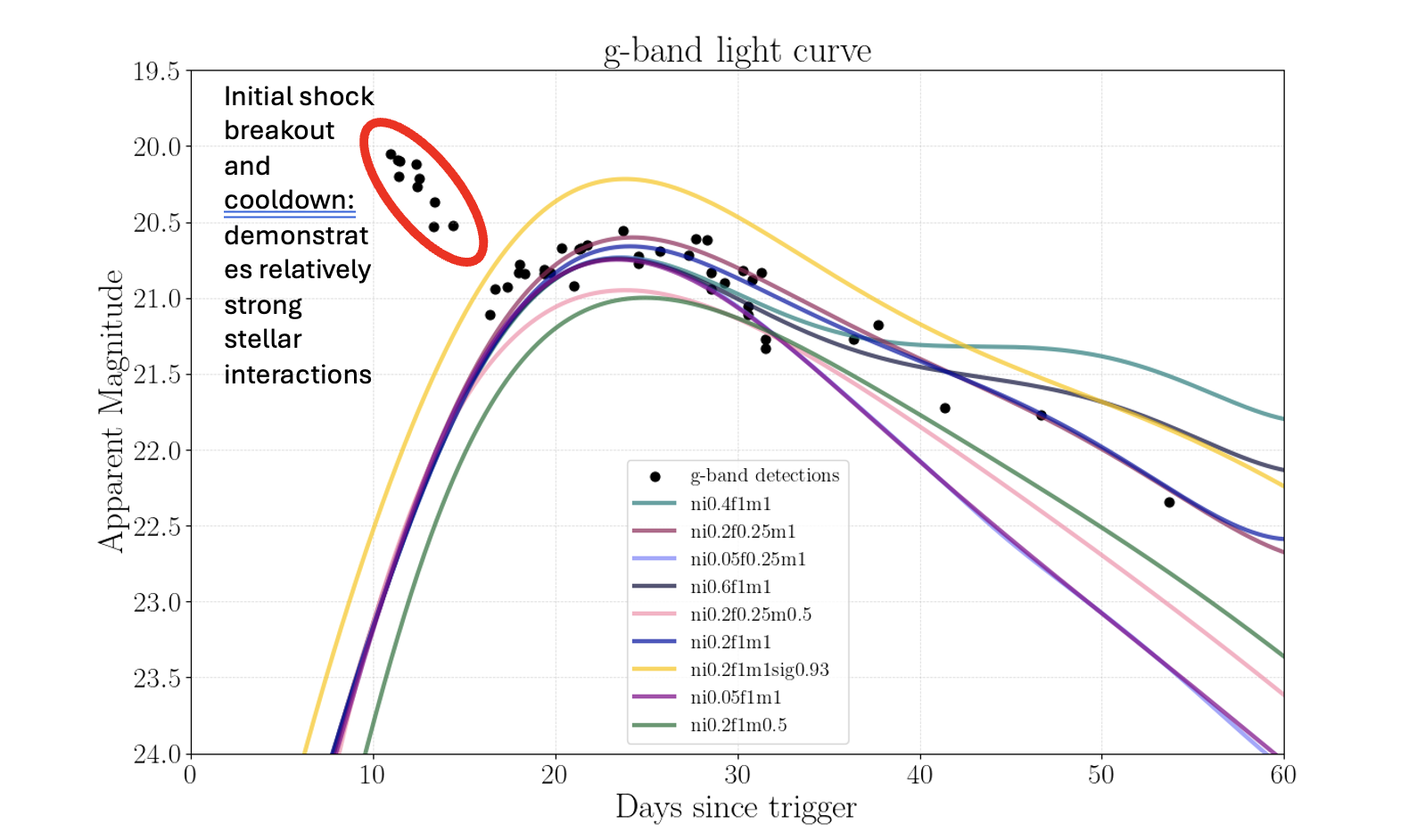}
    \caption{Study of supernova light-curves at peak fitting the Type Ic supernova, 2025kg, g-band emission~\cite{2025ApJ...988L..13R}.  As with many light-curve codes, the light-curve code in these calculations was not designed to capture shock breakout and hence does not fit the initial breakout and cooling phase.  The different models vary the $^{56}$Ni mass, the total mass, the outward mixing of this $^{56}$Ni and the total mass of the ejecta.   This figure includes one model by shock heating.  Although shock heating can explain these features, if we assume that these supernovae are powered by radioactive decay, we can use the light-curve features to constrain the total ejecta mass, the $^{56}$Ni mass and its distribution.  Figure 12 from \cite{2025ApJ...994..259N}} 
    \label{fig:LCcomp}
\end{figure}

Figure~\ref{fig:LCcomp} shows a series of models~\cite{2025ApJ...994..259N} assuming both $^{56}$Ni and shock power sources varying the $^{56}$Ni mass and mixing as well as the ejecta mass.  If there is too much mass, the light-curve will be too bright at late times.  This can be rectified by reducing the mass, increasing the ejecta velocity or, perhaps, by simply increasing the innermost ejecta velocity.  Alternatively, the light-curve can be powered primarily by shock heating, reducing the total $^{56}$Ni mass ejected.  This demonstrates the difficulty in inferring explosion properties from light-curves alone.

\subsection{Transient Spectra}
\label{sec:spectra}

We have discussed a degeneracy between mass and velocity in the duration of the supernova light-curve.  Spectra provide a means to disentangle these effects.  By measuring the spectra during peak, astronomers can probe the composition and velocity of the ejecta at the photosphere~\cite{1994MNRAS.269..764H,2016MNRAS.458.1618D,2025A&A...694A.132D}. Spectra are easiest to obtain at peak luminosity when the number of photons emitted is greatest, but the spectra are much more difficult to interpret.  At this time, the photosphere is still well beyond the stellar center (Figure~\ref{fig:SNLC} and we only measure the velocity distribution (through doppler shifts) and abundances beyond that photosphere.  It can constrain the progenitor, determining the abundances of helium and hydrogen in the star allowing us to type the supernova and constrain its progenitor.

\begin{figure}[ht]
\includegraphics[scale=0.37,angle=0]{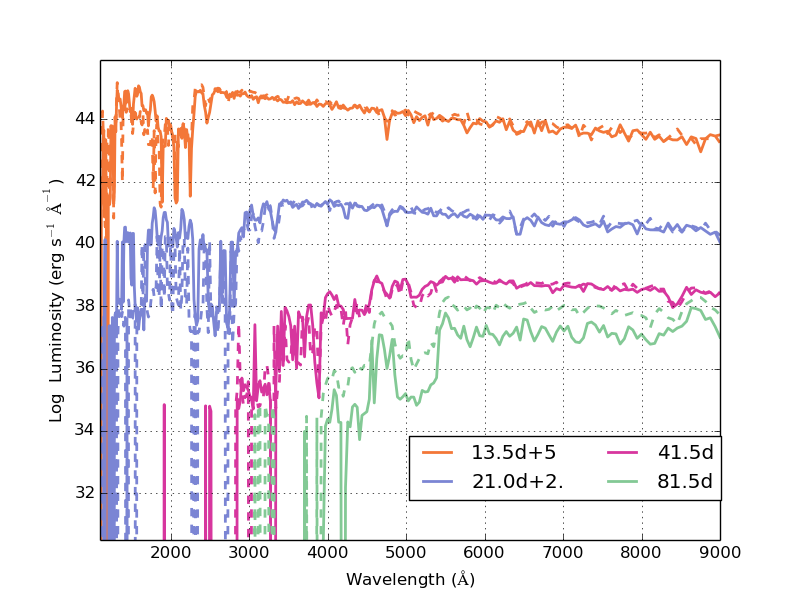}
\includegraphics[scale=0.37,angle=0]{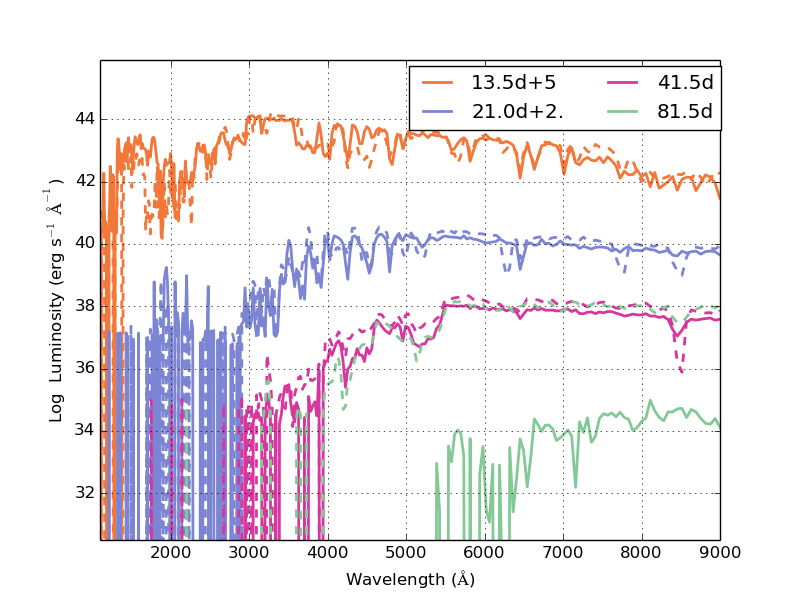}
\caption{Spectral emission (${\rm erg \, s^{-1} \AA^{-1}}$) as a function of wavelength ($\AA$) for our strong (left) and weak (right) explosions of a 15\,M$_\odot$ progenitor.  These models include both low-mix solution with $M_{{\rm mix}}=0.25, f_{{\rm mix}}=1.0$ (solid) and fully mixed with $M_{\rm mix}=13, f_{\rm mix}=1.0$ {\bf (dashed)} solutions.  The primary feature of the fully mixed solutions is the appearance of heavy metal lines in the early-time UV spectra.   At late times, the $^{56}$Ni-heating in the low energy explosion alters the spectra dramatically as it did for the band emission.  However, this radioactive heating can be mimicked by shock heating.  Figure 1 from~\cite{2023ApJ...956...19F}} 
\label{fig:spectra}
\end{figure}

In Figure~\ref{fig:spectra}, we compare simulations~\cite{2023ApJ...956...19F} modeling the spectra from highly-mixed models
and low-mixing models at early times from before peak and throughout the plateau phase.  With high resolution, line broadening will be able to distinguish some features.  The presence of line features from Si-group and Fe-group elements in the early time spectra (primarily in the UV) is a clear indication of extensive mixing.  The large difference between the low-mix and high-mix solutions with our low energy explosion arises because, at these times, $^{56}$Ni heating starts to alter the light-curves in these models (as seen in our band light-curves).  Also note that these comparisons show extreme results (homoegeneously mixed models for our highly-mixed solutions) and the amount of heavy elements in the spectra will be less for any realistic model.

Spectra at late times have proven to be much more powerful probes of the ejecta velocities and abundances~\cite{2017hsn..book..795J}.  At late times, the photosphere moves inward, exposing most of the star (Figure~\ref{fig:latetime}).  The late-time light curves provide constraints on the amount and distribution of radioactive isotopes created in the explosion. The spectral line strengths allows inferrence of ionic masses, emitting volumes, and physical
conditions. Late time spectra have been used to probe asymmetries, abundances, velocity of innermost ejecta~\cite{2022ApJ...928..151F} and provides clues into the energy sources of the explosion~\cite{2022ApJ...928..151F}.  Grids of models provide a means to directly compare to data~\cite{2023A&A...677A...7D}, but this requires obtaining spectra well beyond the supernova peak, observed for only a small fraction of all supernovae.  Obtaining this data will have a much greater impact on our understanding of supernovae than obtaining large numbers of light-curves.

Even with increased data, the analysis of spectra poses a number of problems for astronomers.  Uncertainties in the atomic physics and, more importantly, the out-of-equilibrium physics dictating the population of the atomic level states, can lead to very different interpretations of a spectral line feature.  The inferred abundances from the same spectra have varied by orders of magnitude.  As we shall discuss in Section~\ref{sec:spectheory}, considerable theory work, bringing together atomic and plasma physicists (theory and experiment), numerical physicists, and astronomers is needed to improve the power of this diagnostic.

\subsection{Transient Gamma-Rays}
\label{sec:g-ray}

\begin{figure}
    \includegraphics[width=0.9\textwidth]{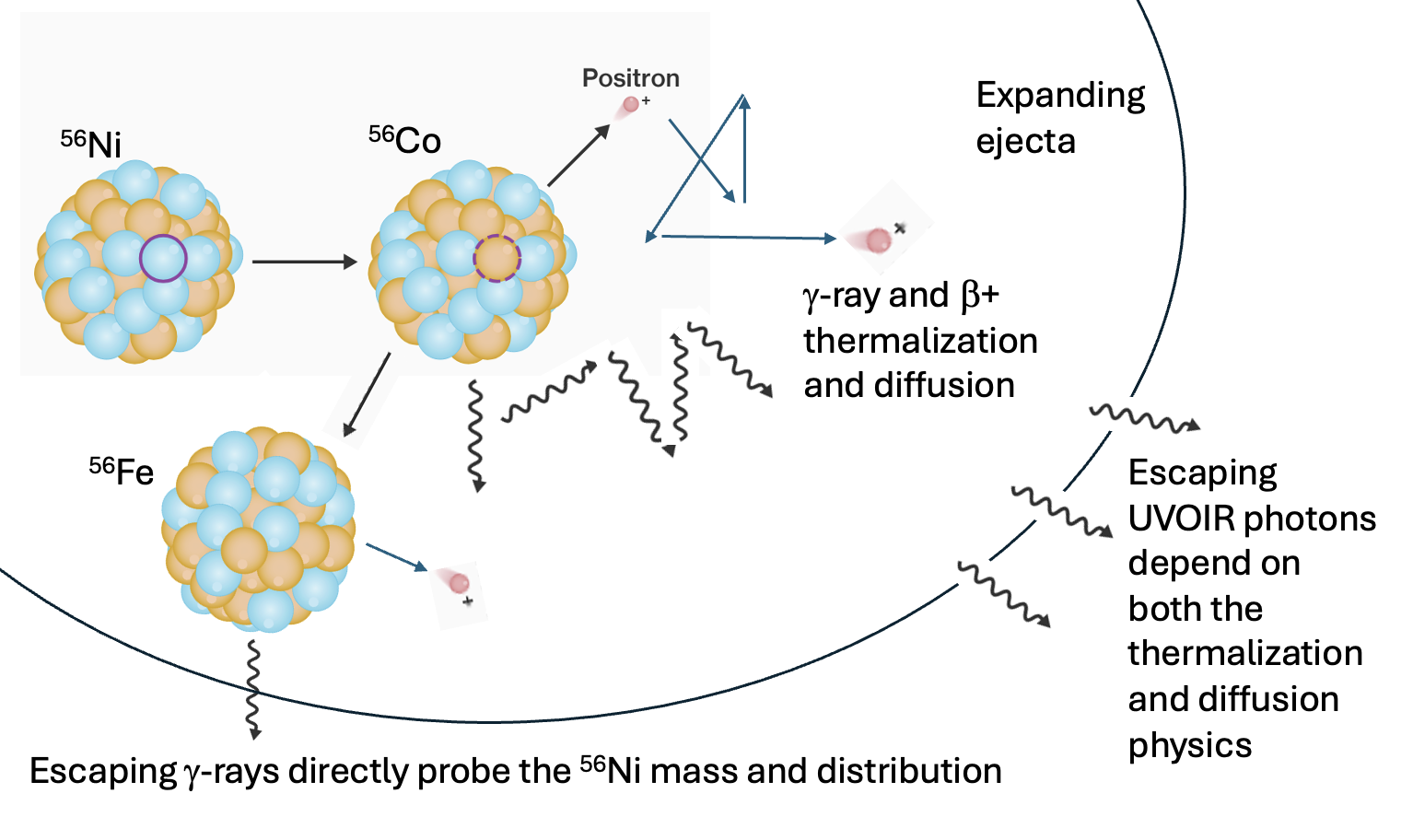}
    \caption{Diagram show one of the key advantages of using $\gamma-$rays to probe the supernova explosions instead of UVOIR emission.  $\gamma-$rays observe the distribution of the $^{56}$Ni decay directly whereas UVOIR emission entirely relies on the thermalization and outward propagation of the radioactive decay products.  This thermalization makes it difficult to distinguish central engine from radioactive decay energy sources.  Figure 5 from \cite{2026arXiv260104464F}.} 
    \label{fig:gammadiag}
\end{figure}

One way to avoid many of the complications facing UVOIR light-curves and spectra is to observe directly the source of energy through the $\gamma-$rays emitted in radioactive decay.  $^{56}$Ni is the primary isotope produced in the extreme conditions of core-collapse supernovae.  The decay of $^{56}$Ni and its daughter product $^{56}$Co emit both gamma-rays and positrons that can either be observed directly or by depositing energy into the ejecta, down-scattered to photon energy in the ultraviolet, optical and infra-red energy (UVOIR) bands.  $\gamma-$ray telescopes observe the decay photons directly (Figure~\ref{fig:gammadiag}), avoiding the detailed modeling required to capture the down-scattering, re-emission and atomic level populations and line-opacity implementations(recall Sections ~\ref{sec:LC} and ~\ref{sec:spectra}).

To date, gamma-ray observations have provided the most exacting constraints on the supernova engine.  The gamma-rays from SN 1987A led to the initial argument for mixing in the supernova engine.  The early rise of the gamma-ray emission required that the $^{56}$Ni produced in the engine must somehow be mixed out into the outer layers of the supernova ejecta~\cite{1988ApJ...329..820P}.  This drove modelers to study asymmetries in the progenitor~\cite{1998ApJ...496..316B} and the engine~\cite{1994ApJ...435..339H}.   The Doppler shift in the $^{56}$Ni decay lines and, at later times, the iron lines, ultimately placed strong constraints on the distribution of this $^{56}$Ni~\cite{2003ApJ...594..390H,2005ApJ...635..487H}. 

\begin{figure}[ht]
\includegraphics[scale=0.38,angle=0]{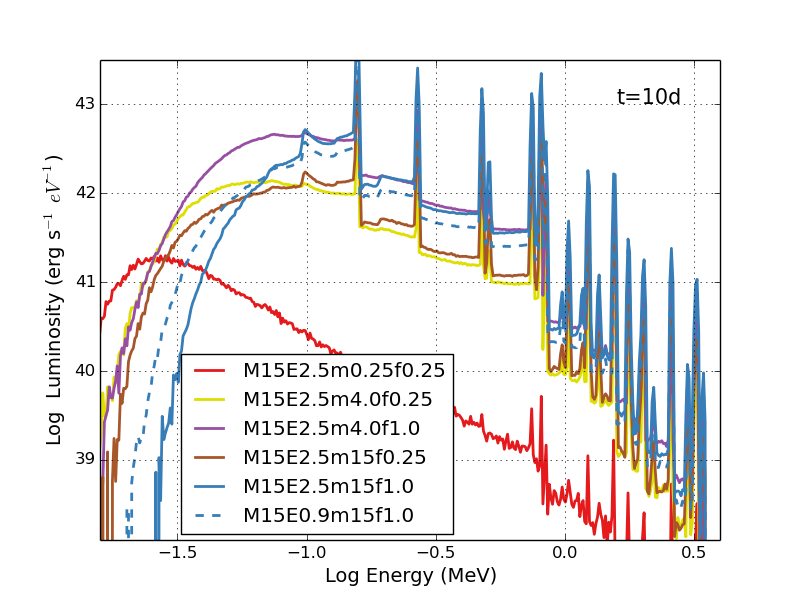}
\includegraphics[scale=0.38,angle=0]{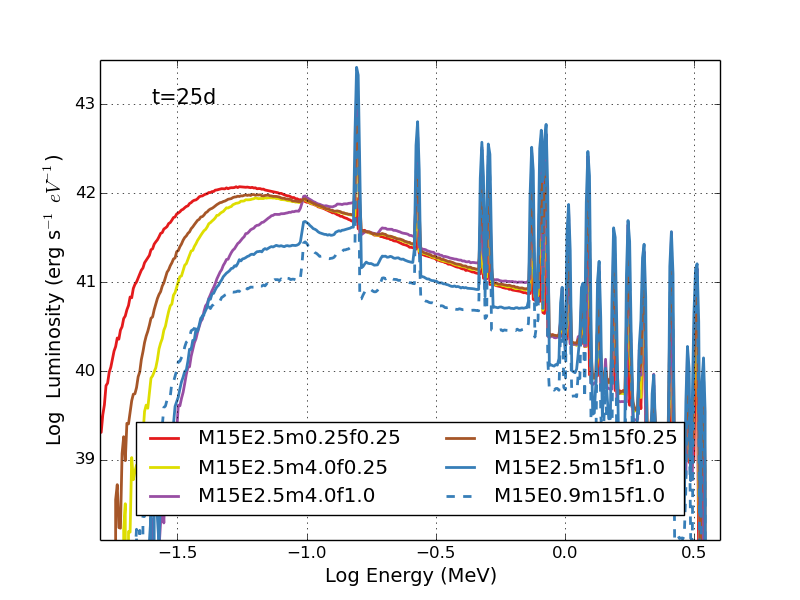}
\includegraphics[scale=0.38,angle=0]{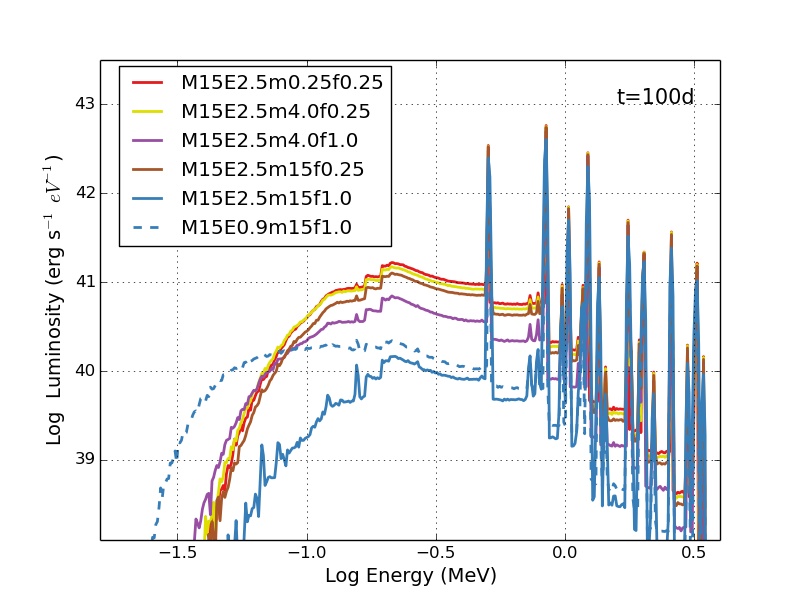}
\includegraphics[scale=0.38,angle=0]{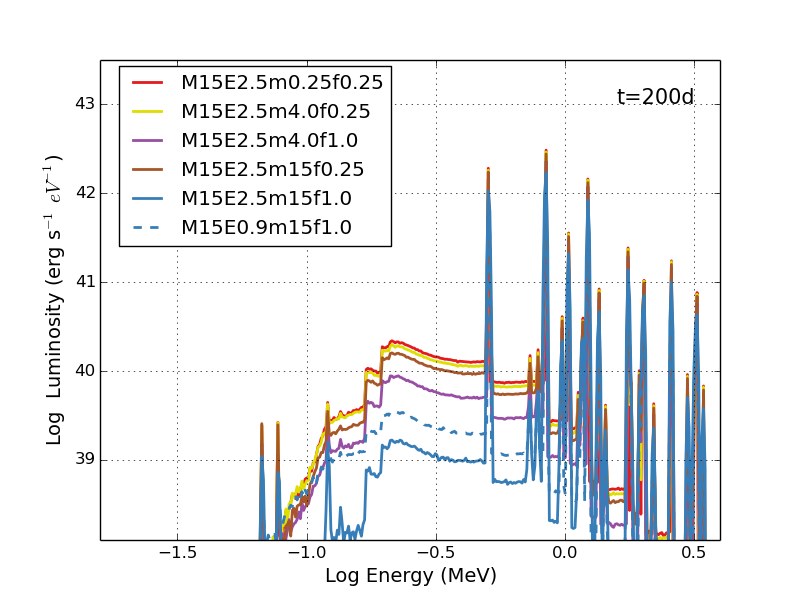}
\caption{X-ray and gamma-ray spectra for 5 mixed explosion models~\cite{2023ApJ...956...19F}:  4 models of our strong 15\,M$_\odot$ explosion with different mixing properties and one weak explosion.  The shape of the continuum and its evolution allows us to differentiate the different models.  Figure 2 from \cite{2023ApJ...956...19F}.} 
\label{fig:grayspectra}
\end{figure}

The outward mixing of $^{56}$Ni first led astronomers to the convection-enhanced neutrino-driven mechanism and remains a strong probe of this mixing.  
Figure~\ref{fig:grayspectra} shows the $\gamma-$ray spectra at four different times from five models with a fixed $^{56}$Ni mass, but varying the amount and extent of the mixing of the $^{56}$Ni of a 15\,M$_{\odot}$ progenitor star~\cite{2023ApJ...956...19F}.  Even with the same $^{56}$Ni yield, the continuum and line flux varies by a factor of 10.  For the models with extended mixing, the line features at early times will be strong.  The models with low mixing only produce such strong lines at late times when the $^{56}$Ni in their explosions can escape.  The low-energy continuum emission, arising for many scatterings of the MeV photons, quickly becomes strongest for the models with little mixing (where the gamma-rays must scatter, losing energy, many times before escaping).

The efficacy of individual spectra to differentiate models depends on observation epoch and resolution of the acquired spectra (i.e., Figure~\ref{fig:grayspectra}). Line intensities and their profiles can be characterized using high-resolution gamma-ray spectroscopy; sensitivity limitations require a Galactic supernova. In addition, detailed line-based population studies are precluded by the low SN event rate, at least for the current generation of space-based telescopes (e.g., \cite{2021arXiv210910403T}).

\begin{figure}[ht]
\includegraphics[scale=0.38,angle=0]{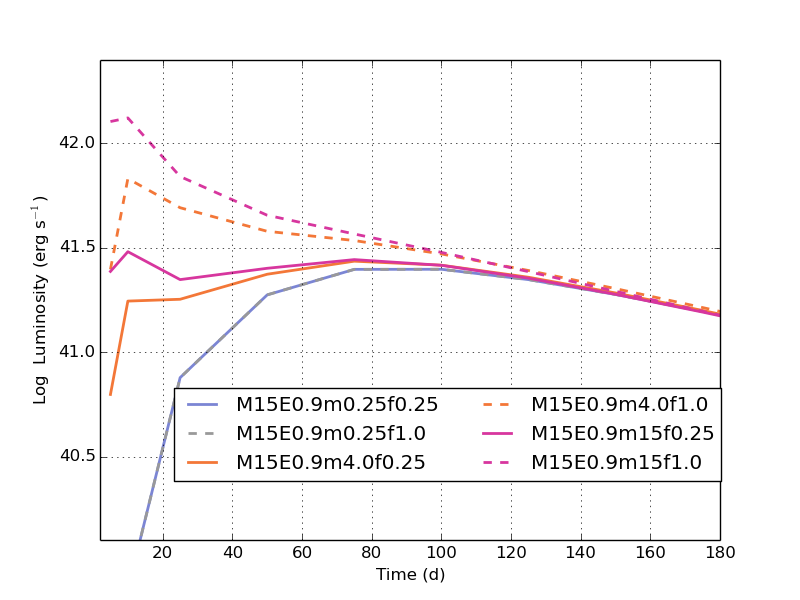}
\includegraphics[scale=0.38,angle=0]{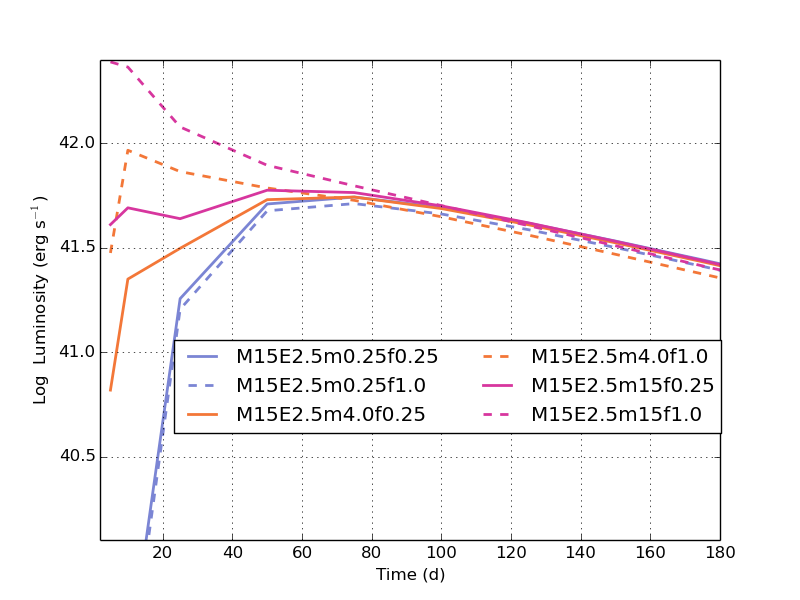}
\caption{Gamma-ray luminosity (photons above 100keV) as a function of time for our suite of 15\,M$_\odot$ explosions.  The gamma-ray lightcurve is sensitive to both the extent and amount of mixing.  The rise time depends on the extent and the luminosity depends more sensitively on the amount of mixing.  Late time signals are sensitive to the total $^{56}$Ni yield.} 
\label{fig:graylc}
\end{figure}

Time evolution is a valuable alternative diagnostic tool. Figure~\ref{fig:graylc} shows the gamma-ray luminosity (photon energies above 100\,keV) as a function of time for our different 15\,M$_\odot$ explosions. The interaction of nuclear line photons with the ejecta leads to a gamma-ray continuum that is present at all epochs. Here, all electrons serve as scattering targets at the energies of interest, independent of the thermal environment or ionization states that affect emergent spectra in the X-ray and UV-optical-infrared (UVOIR) bands. The magnitude and shape of the continuum, as well as the nuclear lines, are time-dependent.

The early-time emission, particularly in the first 20--40 days, is sensitive to the extent and amount of the mixing. With early observations the extent of the mixing can be determined from the light curve rise time, or the early-time luminosity if the source distance is known. At late epochs gamma rays can escape freely as the ejecta becomes optically thin; in this late regime the light curves are dominated by nuclear lines and the intensity reflects the amount of $^{56}$Co (and hence $^{56}$Ni) produced. Thus, light-curves spanning extended post-explosion epochs are excellent diagnostic probes of mixing and elemental abundance yields; monitoring the spectral and temporal evolution of gamma-ray emission (both lines and continuum), from early through late post-explosion epochs, contributes to a complete picture that can facilitate model differentiation.

\subsection{Ejecta Remnants}
\label{sec:SNR}

Supernovae eject most of the progenitor star into their host galaxy, producing remnants that are a powerful probes of the engine.  The expansion of these remnants allow astronomers to spatially resolve the remnant, providing a unique window not only into the ejecta composition but also the distribution (and asymmetries) of this ejecta.  Combining a broad set of observations from radio to gamma ray, scientists have been able to probe supernova explosion properties including the explosion energies and asymmetries.  Supernova remnants have provided some of the most direct support for the convective engine.  The difficulties that current 1-dimensional models have in explaining the yields in supernova remnants demonstrates the power of remnant observations to constrain properties of the supernova and its progenitor~\cite{2019MNRAS.489.4444B,2023MNRAS.525.6257B}.  Most of the remnant observations are limited to nearby events, capping the total number of supernovae that can be spatially resolved and mapped out in detail.  

Even with this limited set of well-resolved remnants, this diagnostic suffers from a few complicating factors.    As the blastwave propagates through the material surrounding the star, it decelerates, sending a reverse shock through the supernova ejecta.   Most observations of the supernova ejecta are limited to material heated by this reverse shock and, in many supernova remnants, we are only observing part of the ejecta.  In addition, the circumstellar medium, much of which is driven by mass outflow from the star (from winds or binary interactions) can often deviate from spherical symmetry.  These asphericities in the circumstellar medium can drastically shape the features and instabilities of the outflow~\cite{1996ApJ...472..257B}.  This emission depends on the characteristics of both the supernova explosion and the circumstellar medium~\cite{2013arXiv1305.4137E}.  We can not take for granted that the features in the supernova remnant are caused by asymmetries in the supernova explosion.  Studies of the Cassiopeia A are a prime example of these difficulties~\cite{2004ApJ...615L.117H}.  These studies focused on the shock-heated silicon feature arguing for evidence of a jet-driven explosion.  But by mapping the distribution of $^{44}$Ti, which is produced in the central engine, scientists found a much different picture for the explosive engine, suggesting a multi-lobe asymmetry in the explosions, as predicted by the convective engine~\cite{2014Natur.506..339G,2017ApJ...834...19G}.  After observing the $^{44}$Ti features of Cassiopeia A~\cite{2014Natur.506..339G}, it became clear that the jet-like structures~\cite{2004ApJ...615L.117H} in that remnant were not caused by asymmetries in the supernova explosion (e.g. jet) but perhaps were produced by explosion/circumstellar medium interactions.

The initial conditions are not the only difficulty in using remnants to measure the abundances and abundance distributions from supernovae (and hence constrain the engine properties).  As the supernova blast wave moves outward, many equilibrium assumptions begin to break down.  For example, remnant shocks are typically in the ``collisionless regime" where the ion and electron distributions can deviate from a thermal Maxwellian, e.g.~\cite{1979ApJ...234L.195P,1981AdSpR...1m..71C,1983PhDT.........2N,2022ApJ...929....7V}.  In addition, the atomic level states, typically set by collisional and radiative processes can deviate from equilibrium and even steady-state solutions~\cite{2010ApJ...725.1476P,2014PhDT.......477B,2015A&A...579A..13V,2017ApJ...851...12R,2020ApJ...903....2R,2021MNRAS.504..583S}.  Inferring abundances in such conditions can be challenging.  All of these deviations from equilibrium, including cosmic ray production, are actively studied by the remnant community~\cite{2020ApJ...903....2R,2021NJPh...23e3010Y,2021MNRAS.504..583S,2022ApJ...929....7V}.  

Another complicating factor lies in the fact that although spectral identifications are ideally suited to determine elemental distributions, they are less powerful at differentiating isotopes.  Many elements synthesized in the central engine exist as stable elements in the star.  At solar metallicity, these stable elements can dominate the signal.  For example, in the \cite{1995ApJS..101..181W} yield database, 15-25\% of the iron in solar-metallicity explosions is not produced in the star or the explosion, but instead was there pre-explosion from the solar distribution of the gas forming the star.  In the \cite{2020ApJ...890...35A} models that consist of a much broader range of explosion energies and a more realistic implementation of the explosion to include fallback, this fraction ranges from 8-100\%, where the 100\% models corresponds to a supernova where all of the synthesized iron falls back onto the neutron star.  To unambiguously probe the supernova engine, we must focus on elements which are only present or predominantly present in the innermost ejecta.  

As discussed above, observations of the $\gamma-$rays produced in the decay of $^{44}$Ti decay have proven to be a much more direct observation of the supernova asymmetry and the Cassiopeia A supernova remnant is the prime example of its power.  Unfortunately, such $\gamma$-ray line observations are limited to two core-collapse supernova remnants:  Cassioepia A and SN 1987A.  Both $^{56}$Ni and $^{44}$Ti have been observed in Supernova 1987A~\cite{Tueller1990_ObseGammLineProfSN1987,Boggs2015_44TiGrayEmisLineSN19ReveAsym}.  Both exhibited redshifted features arguing that the explosion has a large fraction of ejecta moving away from us, supporting a strong sinlge-lobed explosion~\cite{2005ApJ...635..487H} from convection with mild rotation or a disk/jet-driven explosion that sent ejecta outward in a single direction.

\begin{figure}[ht]
\centering
\includegraphics[scale=0.9,angle=0]{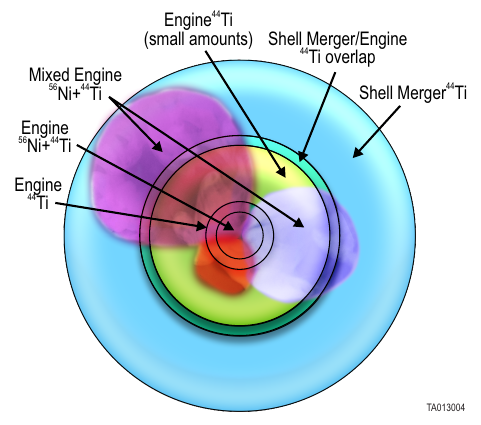}
\caption{Diagram of $^{44}$Ti distribution based on different sources:  neutrino-driven explosions, with and without mixing and shell merger models.} 
\label{fig:tidist}
\end{figure}

$^{44}$Ti exemplifies the power of what we can learn from ejecta remnants.  A number of $^{44}$Ti production sites exist:
\begin{itemize}
    \item Neutrino-driven supernova engine:  $^{44}$Ti is produced in the convective engine and the shocked material just above this engine with tempratures above $4\times10^9 \, K$ and densities from $10^5-19^9 {\rm \, g \, cm^{-3}}$.  The production is sensitive to the details of the shock~\cite{2010ApJS..191...66M}.  Asymmetric explosions match both the total production and multi-modal morphology of observed supernovae~\cite{young_etal_2006,ono_etal_2013,vance_etal_2020}.  If this is the dominant source, we expect the $^{44}$Ti to be slightly more extended than the $^{56}$Ni distribution (Figure~\ref{fig:tidist}).
    \item Jet/Disk engine:  Disk winds can produce considerable $^{44}$Ti at $2\times10^9 \, K$~\cite{2010ApJS..191...66M}.  At these temperatures, $^{44}$Ti is produced but not destroyed to make heavier elements.  The total production fraction in this scenario can be over 1\%.  In this scenario, the $^{44}$Ti distribution will be more centrally located than the $^{56}$Ni.
    \item Shell Mergers:  One of the key issues in stellar evolution is convection and the potential for different burning layers to mix and then merge~\cite{2026arXiv260324758B}.  With this production site, the $^{44}$Ti distribution will be much more extended than the $^{56}$Ni.
\end{itemize}
Current observations of Cassiopeia A~\cite{2014Natur.506..339G} and SN 1987A~\cite{Boggs2015_44TiGrayEmisLineSN19ReveAsym} limit the total contribution of shell mergers~\cite{2026arXiv260324758B}.  But much more work is needed to place stronger constraints.

Remnants can be a powerful tool, but unless we focus on those remnants close enough to observe $^{44}$Ti decay, much more work must be done to reliably infer explosion properties from this diagnostic.  For more discussion on remnants, see~\cite{2023ApJ...956...19F}.

\subsection{Compact Remnants}
\label{sec:compactremnants}

Compact remnant properties are observed in a wide range of objects from pulsars, X-ray binaries and in gravitational waves through mergers.  By measuring the masses of compact stars (neutron stars, black holes), astronomers place constraints on the supernova engine.  {\bf For example, comparing observations to theory predictions based on the properties of the engine models can constrain the growth time of convection in the convection-enhanced neutrino-driven mechanism~\cite{2021ARep...65..937F}.}  In addition, the observed neutron star masses can help constrain the equation of state for dense matter, e.g.~\cite{2012ARNPS..62..485L}.

Traditionally, the information gained has been dominated by observations of X-ray binaries and binary pulsars where emission from these binaries (e.g. radio pulses) probed binary parameters and, hence, placed constraints on the remnant masses.   {\bf Initial mass distribution predictions from these systems were incorrect, arguing for delta-function mass distributions~\cite{1999ApJ...512..288T}.  These observations disagreed with the mass distributions predicted by the convection-enhanced supernova paradigm~\cite{2001ApJ...554..548F}.}  Further observations and improved analyses ultimately led to a better agreement with models. 
There have been several reviews of these distributions (e.g.\cite{2012ARNPS..62..485L}) and an excellent database is at the ``stellarcollapse'' website:  \url{https://stellarcollapse.org/index.html}.

But gravitational waves have opened up a new window into mass measurements~\cite{2021ARep...65..937F}.  Although gravitational waves measure the chirp mass $\mu$ of the binary extremely accurately:
\begin{equation}
    \mu = \frac{(M_1 M_2)^{3/5}}{(M_1+M_2)^{1/5}}
\end{equation}
where $M_1$ and $M_2$ are the masses of the two compact stars (neutron star or black hole) in the binary.  For a given measurement of $\mu$, a range of correlated component masses.  Although this limits what we can learn from these masses, as the number of observed mergers grows, these observations are beginning to provide a picture of the compact remnant mass distribution.  

The Gaia satellite is providing a new window into compact remnants, studying those black holes in wide binary systems~\cite{2024A&A...686L...2G}.  To get accurate orbits, Gaia observations of these wide binaries require data collection over long timescales.  Although only a few black hole systems have been observed, they represent an entirely different population than past observations.  {\bf Another window, providing insight into non-binary black holes is gravitational micro-lensing, a method that is already showing promise~\cite{2022ApJ...933...83S,2022ApJ...933...83S}.}

One of the key observations that could potentially provide real insight into the supernova engine is the existence, and extent, of the mass gap between neutron star and black hole formation ($\sim 2-4\,M_\odot$).  The distribution of masses correlates strongly with the supernova energies and, under the convective engine model, the growth of the convection and the timing of the explosion~\cite{2012ApJ...749...91F,2022ApJ...931...94F}.  But the existence of this mass gap remains a matter of intense study~\cite{2025arXiv250709099R}.  Determining this will help determine the amount of fallback in supernovae and, ultimately, the properties of the supernova engine.

Bear in mind that any of these observations (pulsars, X-ray binaries, {\it LIGO} and {\it Gaia} events) all study different populations with different formation scenarios.  As such, they each represent a subset of the true distribution of compact remnants.  For example, compact binary mergers observed by gravitational waves are produced only in tight binaries that ultimately will merge.  These may only form in particular scenarios of common envelope evolution.  We also do not know the metallicity of the stars that form the {\it LIGO}-observed binaries (we don't know when they formed, just when they merged).  Any metallicity effects will be difficult to determine.  This may be fixed with next generation detectors that can detect binaries out to high redshifts.  On the other hand, the wide binaries observed by {\it Gaia} also likely form in only specific scenarios where kicks are minimal (large kicks would unbind the binary).  To fully understand these binaries, we will need detailed population synthesis models~\ref{sec:popsyn}.

\begin{figure}[ht]
\includegraphics[scale=0.24,angle=0]{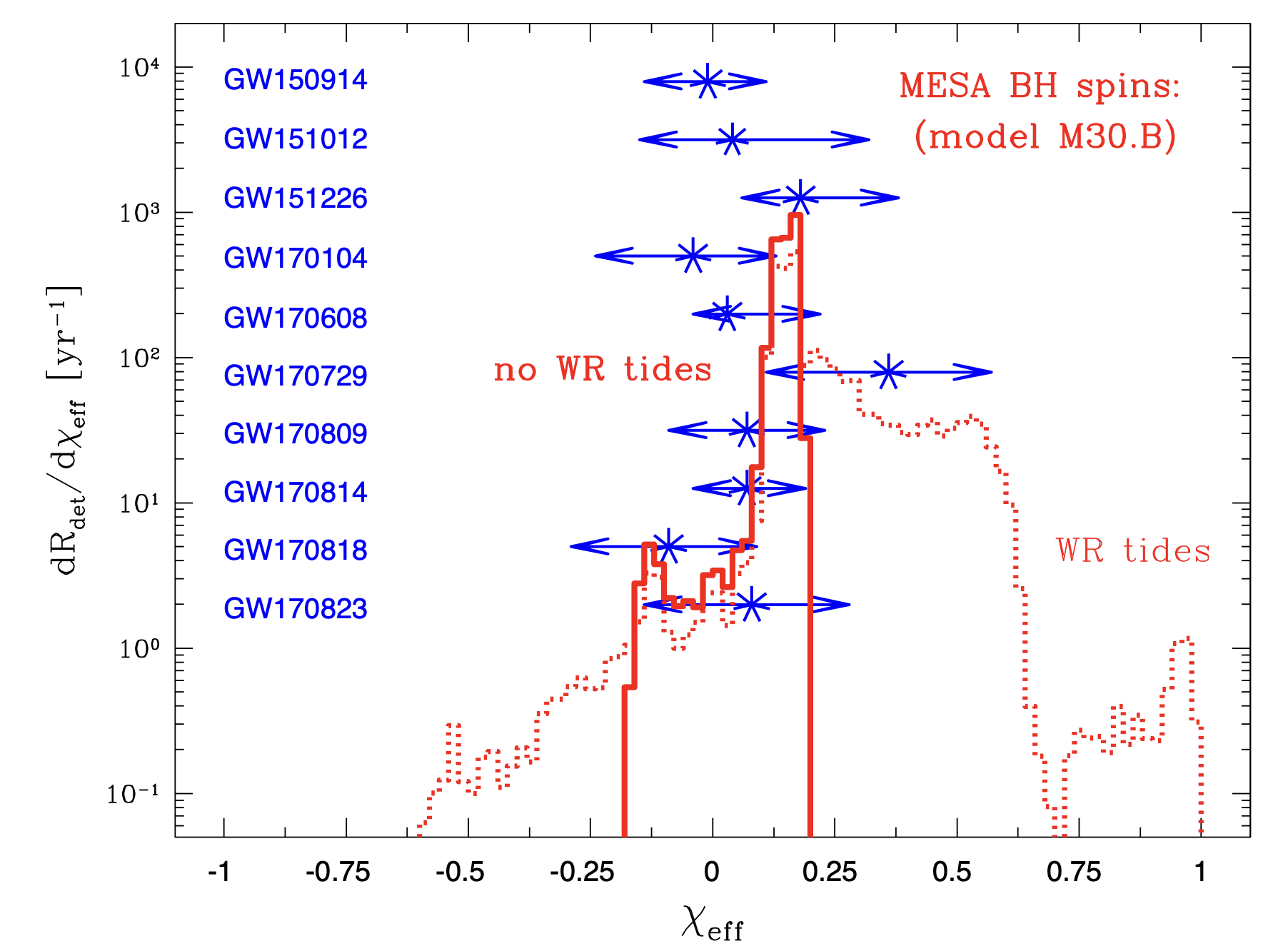}
\includegraphics[scale=0.25,angle=0]{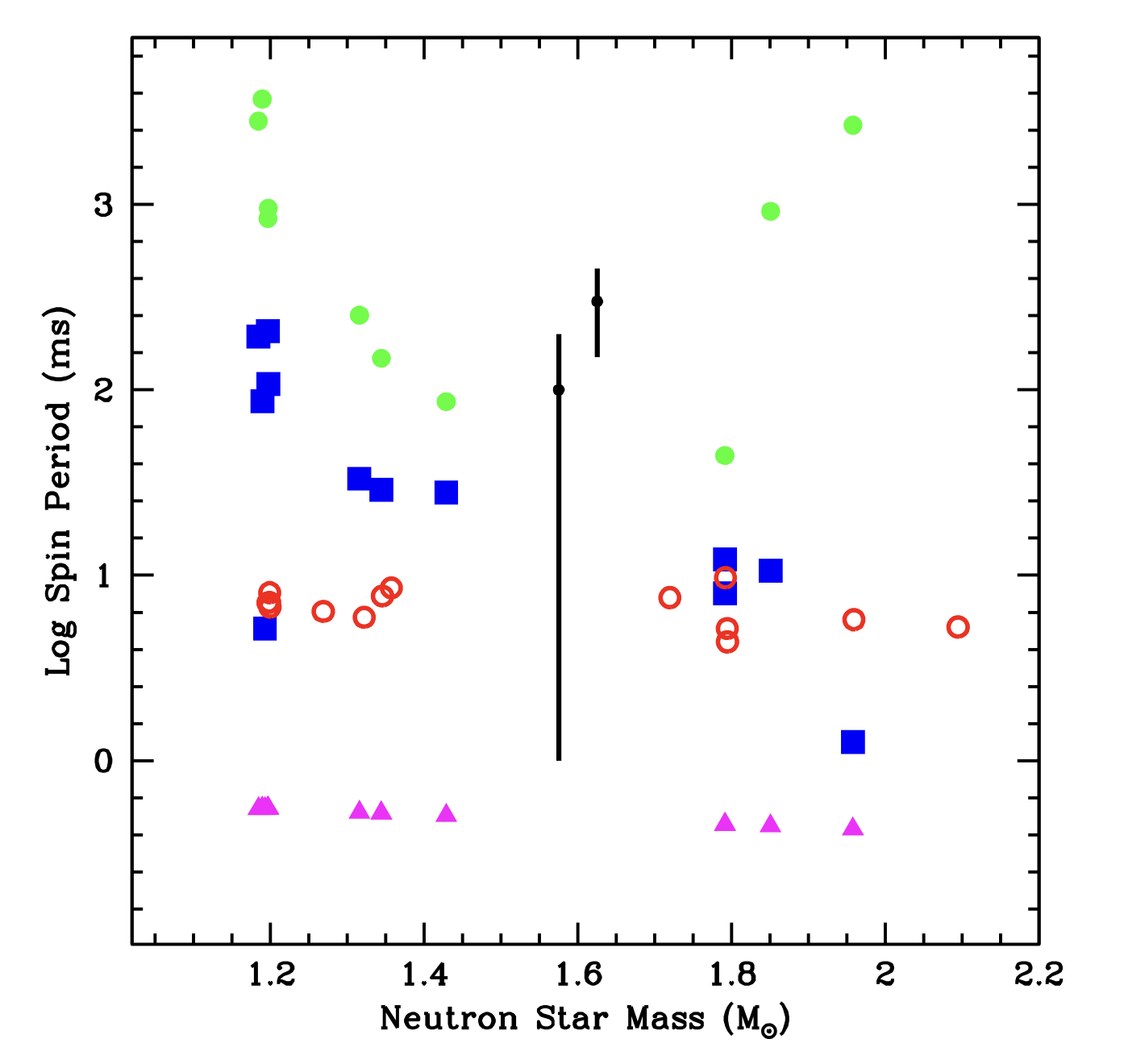}
\caption{Left: Number of merging black hole binaries $\chi_{\rm eff}$ Figure 19 and Figure A.3 from \cite{2020A&A...636A.104B}.} 
\label{fig:crspin}
\end{figure}

Birth spins of neutron stars and black holes place some of the strongest constraints on the nature of the explosion.  Jet/disk-driven or magnetar explosions require extremely high spin periods.  Studying remnant spins requires extracting the effects of binary evolution and mass accretion from the observations.  A detailed study of binary population synthesis results to these observations has already been done~\cite{2020A&A...636A.104B} and we summarize it here.  This work found that the observed spin rate suggest moderate to strong angular momentum coupling (for both the neutron star and black hole spins).  Figure~\ref{fig:crspin} shows the comparisons between models and data for both neutron stars and black holes.  Although this does not preclude tidal spin-up, it does argue that such tidal spin-up pathways are rare, occuring for less than 10\% of all core-collapse progenitors.  Combined with our current understanding of stellar evolution, this argues that the jet/disk- or magnetar- engines can account for less than 10\% of all observed supernovae.

\section{Theory Needs}
\label{sec:models}

As we have discussed in this paper, the impact of EM observations relies on complex models that tie models of the progenitor and the engine to the broad range of astrophysical observables.  Here we briefly describe the theoretical studies that must be enhanced to truly take advantage of the upcoming data.

\subsection{Progenitors}
\label{sec:progenitors}

The uncertainties in progenitor models include:
\begin{itemize}
    \item Stellar Mixing:  e.g. shell mergers, angular momentum transport
    \item Stellar Burning:  e.g. $^{12}$C($\alpha$,$\gamma$)$^{16}$O
    \item Radiation Transport:  e.g. detailed opacities, multi-group effects
    \item Stellar Mass Loss:  e.g. Stellar winds, explosive mass loss 
\end{itemize}

As an example of recent studies of stellar progenitors lies in the study of stellar mixing and the presence of shell mergers.  These mergers shape the structure of the star that will alter the size of the core.  This inner core sets the fate of the supernova engine.  But these shell mergers also affect the yields.  There is a temptation to rely on the best 3-dimensional calculation to solve this problem.  But the viscosity in 3-dimensional simulations is much higher than that in nature.  Although these simulations can provide critical insight, they must be combined with analytic understanding.

These models can be tested against observations.  C--O shell mergers have been investigated mostly because of their nucleosynthetic signatures in the production of p-process nuclei beyond Fe~\cite{Ritter2018_NuGrStelDataSetIIStelYiel, Roberti2023_GproNuclCoreSupeNoveAnalGpro} and odd-Z nuclei~\cite{Issa2025_3DMacrPhysLighOddzElemProd, Roberti2025_OccuImpaCarbShelMergMassStar, Issa2026_Impa3DMacrNuclPhysP-nuOC}. In light of these studies, the XRISM collaboration~\cite{Audard2025_ChloPotaEnriCassSupeRemn,Sato2025_InhoStelMixiFinaHourCassSupe} has recently analyzed the emission from Cl, K, Ne, Mg, Si, and Ar in Cas~A, finding many similarities with the nucleosynthetic signatures of C-O shell merger progenitors. Studies have also focused on the production of $^{44}$Ti production~\cite{Issa2026_Impa3DMacrNuclPhysP-nuOC}.

\subsection{Supernova Engines}
\label{sec:snengines}

Other articles in this issue will cover the broad physics that must be studied to understand the supernova engine.  However, as with the stellar model studies, it is critical that these studies must couple the results with an analytic understanding.  To compare to observations, we need to understand the possible range of results.  We stress that detailed simulations are essentially useless without this broader understanding.

\subsection{Nucleosynthetic Yields}
\label{sec:modelnuc}

\begin{figure}
\centering
\includegraphics[width=0.9\textwidth]{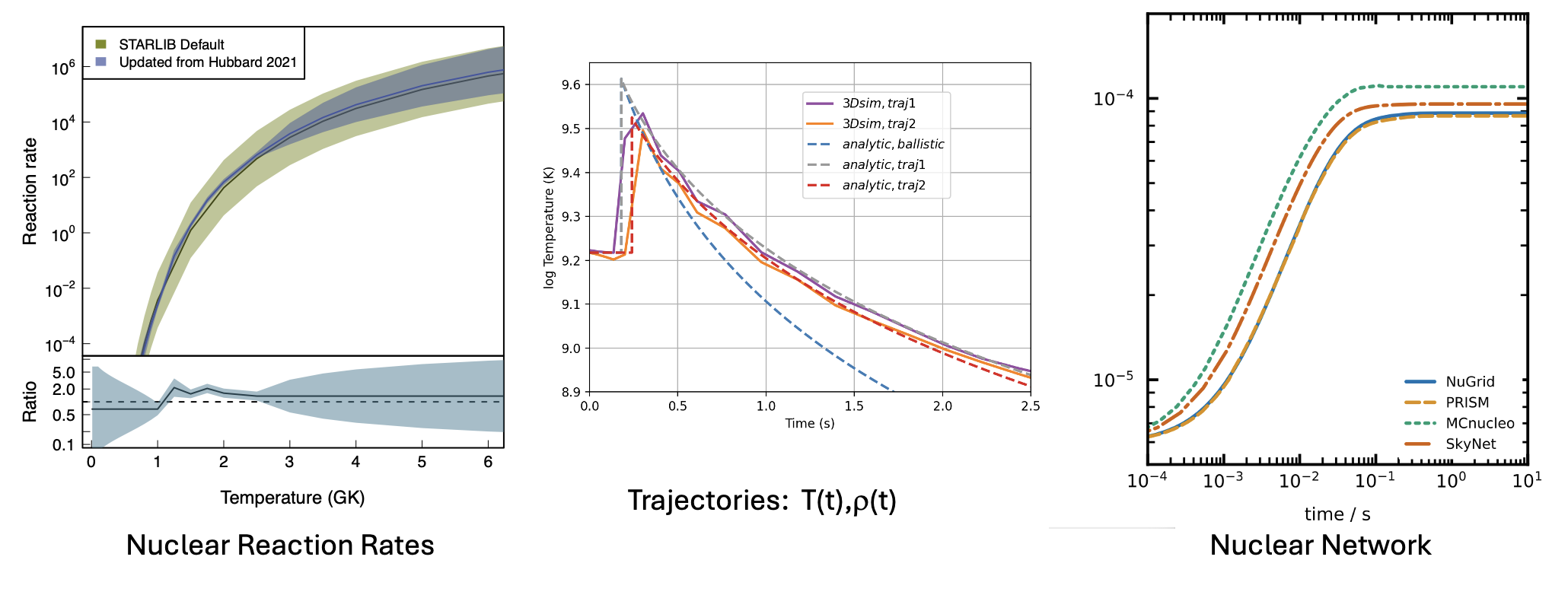}
\caption{The three primary calculations whose uncertainties must be understood to use gamma-rays to probe the supernova engine.  Left panel:  nuclear reaction rates are determined combining experimental data and approximate nuclear physics models that are tuned to fit the experimental data.  The middle panel shows analytic fits to simulated trajectories.  These models must be modified to include the deceleration many of the ejecta experiences.  Right:  Using reaction rates and estimated trajectories, nuclear networks calculate the yiedls.  Uncertainties in the nuclear networks can be minimized through detailed comparisons.}
\label{fig:theory_snnuc}
\end{figure}

The nuclear yields are critical to nearly all of the EM observations, particularly light-curves, spectra, $\gamma-$ray transients and ejecta remnants.  Nucleosynthetic yields produced near the central engine (either in the engine or in the silicon shell just above it) can be a powerful probe of the central engine, but require a broad range of physics.  A recent review provides detailed discussion of the uncertainties in these calculations~\cite{2026arXiv260104464F}.  This review divided the uncertainties into 3 different categories:  progenitor studies, trajectories, nuclear-physics models.  We have discussed the progenitor studies above (Section~\ref{sec:progenitors}).  Here we simply list the uncertainties in trajectories and nuclear physics.  Figure~\ref{fig:theory_snnuc} shows the range of theory needed to produce detailed yields.

The uncertainties in the trajectories include:
\begin{itemize}
    \item Explosion properties:  e.g. energetics, asymmetries
    \item Electron fraction:  e.g. neutrino effects
    \item Explosion duration:  e.g. long-lived explosion properties
    \item Explosion Propagation:  e.g. detailed shock physics through uncertain stellar conditions
\end{itemize}
Trajectory modeling pushes the limits of our numerical capabilities.  Including the long-lived supernova engine in the required 3-dimensions pushes our numerical capabilities.  Modeling the shock propagation requires a level of resolution that is beyond our multi-dimensional models.  Any solution to these trajectory studies will require combining 1-, 2- and 3-dimensional models that capture the full range of potential explosion properties.  To compare with observations, we must understand the range of potential solutions.  This should be a primary goal of the engine modelers if we are to truly use EM observations to constrain the engine. 

Calculating nuclear physics burning rates from first principles is beyond current numerical capabilities. {\bf A full quantum study of nuclear cross-sections without approximations is beyond theoretical capabilities at this time.}  Any approach at this time requires coupling numerical models with targeted nuclear experiments.  Quantum computers have the potential drastically improve the numerical models, but it is unlikely that this will alter the theory plus experiment approach for many decades, if at all.

\subsection{Supernova Light-Curves and Spectra}
\label{sec:spectheory}

Although there are a large number of supernova light-curve and spectral observations and the number of these observations are growing rapidly, our understanding of these observations limit what we can learn from them.  The modeling of supernova light-curves and spectra requires combining multiple physics fields, each with their own uncertainties, making it difficult to make quantitative analyses.  These fields include:
\begin{itemize}
    \item Atomic Physics:  Current techniques make assumptions that simplify the unsolvable, multi-body Schr\"odinger equation for atoms and molecules by treating electrons independently, freezing nuclear motion, or averaging potentials.  Different groups use different approximations.  Current solutions produce varying results and most groups calibrate off of laboratory experiments.  It is possible that quantum computers will ultimately reduce the uncertainties in some of these assumptions, but considerable work is needed to identify these uncertainties and the exact nature of the quantum computing solution.  
    \item Plasma Physics:  The conditions in supernova explosions are far from equilibrium.  First off, supernova opacities have a complex structure that produce photon spectra that can be very different from a true blackbody.  Even if we can approximate the radiation from the supernova-ejecta photosphere as a blackbody, most of its interactions will be above the photosphere and this can be described as a dilute blackbody at best.  In addition, energetic electrons produced in the decay of radioactive isotopes lead to conditions that are far from equilibrium.  Magnetic fields can play a major role in the transport of these charged particles, further pushing the plasma physics difficulties.  These effects feed into both the atomic physics and radiation-hydrodynamics calculations.
    \item Radiation-Hydrodynamics:  It is becoming increasing evident that shock interactions play an important role in supernova observations.  Although accurate high-order radiation transport models exist (e.g. discrete ordinates, spherical harmonics, Monte Carlo), the coupling of radiation transport with hydrodynamics remains an active area of research.  In many cases, it is impossible to resolve the the mean free path of the radiation in the material and, hence, the uncertainties in the material feedback to radiative heating can drastically change the shock-interaction results.
\end{itemize}

\begin{figure}
    \centering
    \includegraphics[width=5.5in]{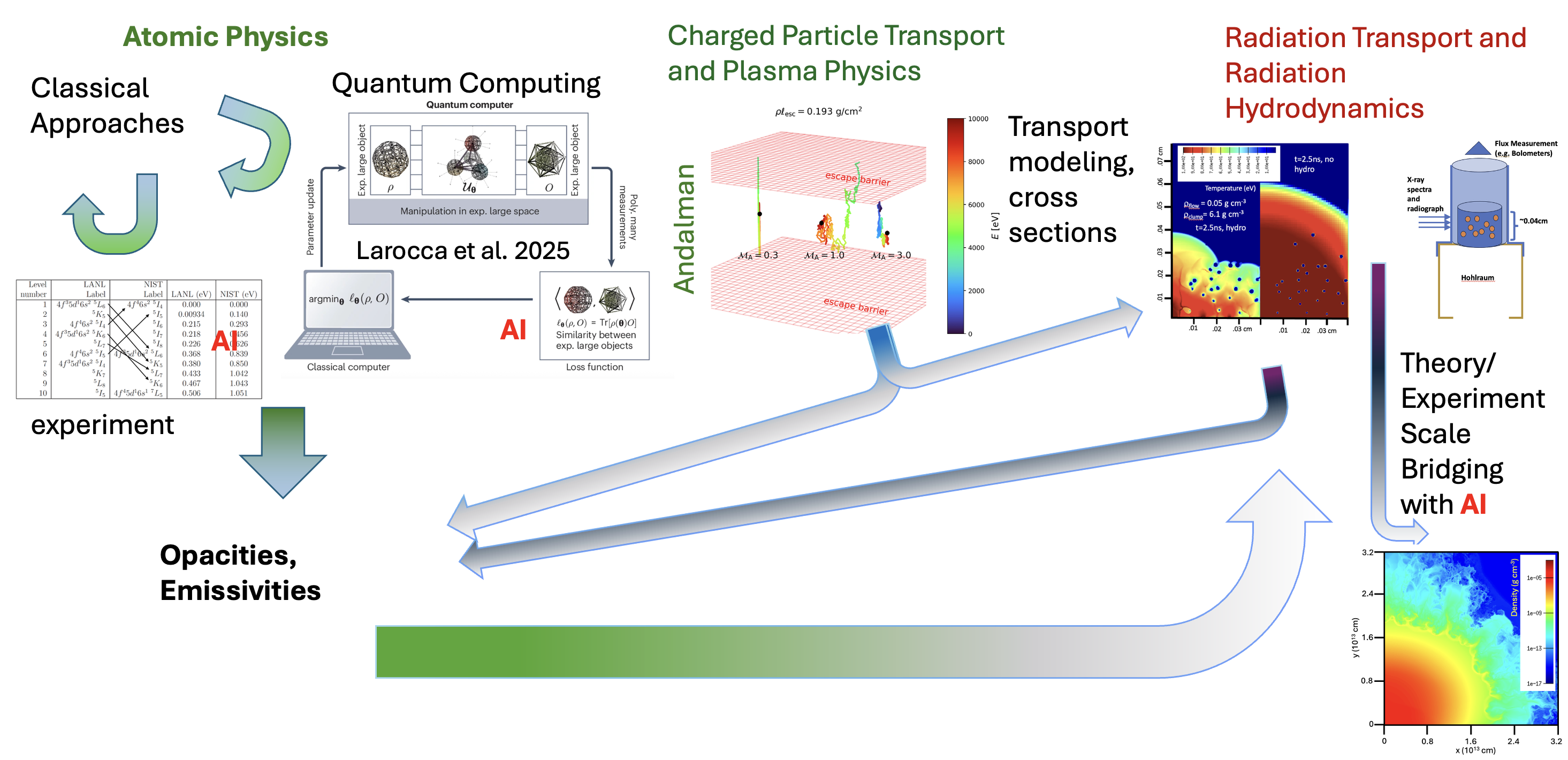}
    \caption{The interplay of physics and physics-studies for supernova light-curves and spectra.}
    \label{fig:modtrans}
\end{figure}

The interplay of these physics disciplines is shown in Figure~\ref{fig:modtrans}.  Shock interactions have become one of the newer pieces of physics affecting light-curves and spectra~\cite{2017hsn..book..403S}.  A growing number of studies have focused on studying the physics behind shock interactions for shock breakout and pre-peak supernova light-curves~\cite{2020ApJ...898..123F}.  Uncertainties in the coupling of radiation to the material place limitations on these results, but a growing number of laboratory experiments have been developed to study this physics.  Shock interactions rely on understanding the circumstellar medium surrounding the supernova progenitor.  Understanding the nature of this circumstellar medium requires improved models of stellar mass loss mechanisms and binary interactions.

{\bf Capturing both detailed opacities and radiation hydrodynamics has proven challenging.  Astrophysicists use a broad range of techniques.  In many cases, astronomers use semi-analytic`(e.g. Arnett Law~\cite{1982ApJ...253..785A}) or simple, diffusive single-temperature transport methods~\cite{2015ApJ...814...63M,2022ApJ...929..177T,2025ApJ...994..259N}.  More detailed transport techniques are required to implement realistic opacities.  Capturing line opacities, particularly in problems where there is considerable Doppler broadening can be challenging and a number of approaches ranging from pure Sobolev solutions\cite{1963trt..book.....S} to expansion opacities have been developed~\cite{1977ApJ...214..161K,1983ApJ...272..259F,1993ApJ...412..731E,1996AstL...22...79B,2003A&A...401...43W}.  Line-dominated opacities are extremely important for stellar winds, and this community has developed specialized opacity implementations and transport techniques to capture this physics~\cite{1980ApJ...241.1131C,2016MNRAS.458.2323K,2018A&A...611A..17S,2019A&A...631A.172D,2021A&A...648A..94L} although more-recent approaches have begun to introduce moment transport techniques such as flux-limited diffusion~\cite{2022A&A...665A..42M,2025A&A...696A.131N} and discretized angle solutions~\cite{2022ApJ...929..156G,2022ApJ...933..164G}.}

Although the uncertainties in supernova transients limit what we can learn from them at this time, coupling the growing set of experimental data, theoretical studies and computational advances to the rapidly increasing set of data may make this diagnostic one of the strongest constraint on the supernova engine in the future.

\subsection{Compact Remnants and Binary Population Synthesis:}
\label{sec:popsyn}

Understanding the nature of the circumstellar shock interactions is not the only field that requires and understanding of binary effects.  Compact remnant masses also rely on a detailed understanding of binary population studies.  Any particular remnant population is likely to be biased by its formation scenario:  e.g. X-ray binaries are limited to systems sufficiently close to have a mass transfer stage (either Roche lobe overflow or wind interactions), merging systems are similarly biased by systems that will merge in a Hubble time.  Merging systems have the additional constraint that we measure the merger redshift, not the formation time of the binary.  To extract the formation as a function of metallicity requires detailed population models.  Population models, in turn, rely on an understanding of a broad range of physical models and physics:
\begin{itemize}
    \item Stellar Evolution:  The key progenitor properties for compact remnants include uncertainties in the stellar winds (mass loss that can affect binary separations), shell mergers (that affect the core and, hence, the remnant properties), angular momentum transport (to determine the role of rotating supernovae), and stellar radii (determining which models go through mass transfer phases)
    \item Supernova Properties:  The key supernova properties are remnant masses, spins and asymmetries that produce kicks{\bf ~\cite{2020A&A...636A.104B,2022ApJ...931...94F,2022MNRAS.516.2252O}.}
    \item Binary Models:  Population synthesis models are critical.  These models need to identify the critical physics and direct detailed studies to improve our understanding of these uncertainties{\bf~\cite{2008ApJ...672..479O,2008ARep...52..299B,2023ApJS..264...45F,2025ApJS..281....3A,2026ApJ..1002..105R}.}
\end{itemize}

\section{Summary}

{\bf Time domain astronomy is about to undergo a revolution with a broad range of new and upcoming facilities beinng developed including the Rubin telescope, Einstein Probe, COmpton Spectrometer and Imager (COSI), Ultraviolet Explorer, next generation gravitational wave telescopes to name a few.  Combining these multi-band and multi-messenger (including gravitational waves and neutrinos) signals will allow scientists to ultimately pinpoint the nature of supernova explosions.}

A broad number of EM observations contribute to our understanding.  These include shock breakout, transient light-curves, spectra and $\gamma-$rays, ejecta and compact remnants.  Combined, they are powerful diagnostics of the supernova engine.  But to leverage these observations, astronomers must bring together a broad set of disparate physics and detailed coupling of different numerical models.  Solving these problems will test the ability of interdisciplinary physicists, mathematicians and computer scientists.

\ack{The work by CLF was supported by the US Department of Energy through the Los Alamos National Laboratory. Los Alamos National Laboratory is operated by Triad National Security, LLC, for the National Nuclear Security Administration of U.S.\ Department of Energy (Contract No.\ 89233218CNA000001).}






\bibliography{refs}{}
\bibliographystyle{iopart-num}

\end{document}